\documentclass[prd,preprint,superscriptaddress,nofootinbib,longbibliography,aps]{revtex4-1}
\usepackage[utf8]{inputenc}
\usepackage{comment}
\usepackage{geometry}
\usepackage{soul}
\usepackage{graphicx}
\usepackage{longtable}
\usepackage{amsmath}
\usepackage{color} 
\usepackage{multirow}
\usepackage{amssymb}
\usepackage{subfigure}
\usepackage{tabularx}
\usepackage{microtype}
 \usepackage[linktoc=all]{hyperref}
 \hypersetup{
    colorlinks=true,
    linkcolor=BlueViolet,
    citecolor=Blue,
    urlcolor=MidnightBlue,
}
\usepackage{cleveref}
\usepackage[table]{xcolor}
\usepackage{stackengine}
\usepackage[normalem]{ulem}
\usepackage{slashed}
\usepackage{color}
\usepackage[dvipsnames]{xcolor} 

\usepackage{bm}
\usepackage{braket}
\usepackage{slashed}
\usepackage{comment}
\definecolor{myblue}{rgb}{ 0.988, 0.078,0.458}

\begin{document}
 
 \title{Forecasting Sensitivity to Modified Dispersion Effects in Pulsar Timing Arrays } 

\author{Jonathan Grée }
\email[]{jonathan.gree@polytechnique.edu}
\affiliation{ \'Ecole polytechnique, Institut polytechnique de Paris, Palaiseau, France}

\author{Qiuyue Liang  }
\email[]{qiuyue.liang@ipmu.jp }

\author{Elisa G. M. Ferreira}
\email[]{elisa.ferreira@ipmu.jp }
\affiliation{Kavli Institute for the Physics and Mathematics of the Universe (WPI), UTIAS, The University of Tokyo, Chiba 277-8583, Japan}

\date{\today}

\begin{abstract}
The pulsar timing array systems have reported a detection of a nanohertz-band stochastic gravitational wave background in our galaxy. It is of interest to use this observation to probe modified gravity and to forecast the sensitivity with which certain deviations can be tested in the coming years. In this paper, we focus on the modified dispersion relation of the tensor modes and its effect on the overlap reduction function of the timing residual cross-correlations. We perform a comprehensive forecast of the phase velocity uncertainty, $\sigma_v$, using a Fisher analysis validated by a mock-data study to account for potential non-Gaussian behavior. We also take into account the sample variance effect and provide an observational timeline for future PTA sensitivity: detecting a $10\%$ or $-1\%$ deviation from the speed of light at the $3\sigma$ level requires $\mathcal{O}(30)$ years of observations. 
\end{abstract}



\maketitle
\tableofcontents

\section{Introduction}
Pulsar timing arrays (PTAs) provide a Galactic-scale detector for nanohertz gravitational waves, making use of the remarkable stability of millisecond pulsar signals. Using many pulsars in the array gives a particular sensitivity to weak common processes that affect the timing of the arriving pulses. Recent PTA data have revealed strong evidence for a stochastic gravitational-wave background (SGWB) in this band. In particular, the NANOGrav 15-year data set shows a common-spectrum red process whose pulsar pair correlations approximately follow the quadrupolar Hellings–Downs pattern expected for an isotropic gravitational wave background~\cite{Agazie_2023,Afzal_2023} in general relativity. Similar evidence for common red-spectrum signals has been observed in the latest European PTA (EPTA)~\cite{EPTA_2023} and Chinese PTA (CPTA) analyzes~\cite{Xu_2023}. Other collaborations such as PPTA, InPTA and MeerKat are focusing on the same detection~\cite{Reardon_2023,Yardley_2011,tarafdar2022indian,miles2023meerkat}. These breakthrough detections suggest the gravitational-wave origin for the nanohertz noise. 

In the PTA analysis, assuming an isotropic stochastic GWB allows us to separate the signal into two parts, the power spectrum and a characteristic angular correlation known as the overlap reduction function (ORF). In general relativity (GR), the ORF is the Hellings–Downs (HD) curve~\cite{Agazie_2023,hellings1983upper} that relates the timing residual correlations to the angular separation between pairs of pulsars. 
In the long-distance limit, the ORF is frequency-independent, and provides unique information of the GWB. 
To date, PTAs have found $3\sim 4\sigma$ evidence~\cite{Agazie_2023,Xu_2023} for the HD correlation compared to a noise-only hypothesis, which still leaves room for alternative correlation patterns. With the advancement of instrumentation, the discovery and precise timing of more pulsars, and the continued extension of observation time, the PTA dataset will achieve progressively higher precision.
 
Beyond confirming a GR background, PTAs create new opportunities to constrain general relativity and test modified gravity theories, which usually contain extra scalar and vector polarization modes~\cite{Gair:2015hra,Qin:2018yhy,Qin:2020hfy,Liang_2021,Bernardo:2023zna} and have modified dispersion relation~\cite{Liang_2023,Bernardo:2023mxc}. 
The ORF offers a clean probe for testing and constraining such modifications. However, vector modes are always decoupled from the source~\cite{Hinterbichler:2011tt} and are therefore suppressed in the production mechanism, whereas scalar modes, although not decoupled, are suppressed in their generation~\cite{deRham:2012fw} and in the solar system by various screening mechanisms (see recent reviews~\cite{Joyce:2014kja, Baker:2019gxo} and references therein ). Consequently, in this work, we focus solely on the tensor modes and their modified dispersion relation. Previous studies~\cite{liang2024testinggravityfrequencydependentoverlap,Liang_2023} have  provided the analytical expression for the modification of the ORF induced by the phase velocity. Current PTA analyzes lack the sensitivity to constrain these effects to a statistically significant level~\cite{Bi:2023ewq,Bernardo:2023pwt,Wang_2024,Wu_2024}. It is therefore of great interest to forecast the PTA sensitivity required to detect deviations from GR at a given confidence level. Previous work has been conducted in harmonic analysis, as discussed in~\cite{Nay:2023pwu,Bi:2026jeu}.

In this work, we use a Fisher forecast on the ORF to relate the uncertainty on the measurement of the gravitational wave phase velocity to the precision of the measurement of the timing residual correlation. Since the true value of the phase velocity is unknown, we treat it as a parameter and perform the analysis using fiducial phase velocities. By approximating the data as Gaussian distributions around our model of the overlap reduction function, the Fisher forecast provides an optimistic estimate of the uncertainty in measuring the phase velocity. This uncertainty is a function of both the fiducial phase velocity and the precision of the correlation measurements between pulsar pairs. We further provide tests to justify these assumptions and translate the uncertainty into observation time by assuming a specific rate of new millisecond pulsar discoveries. The sample variance effect on the overlap reduction function is properly considered.

This paper is organized as follows. Sec.~\ref{sec1} describes the peculiarities of the PTA analysis in the framework of a modified dispersion relation, derives an analytic expression of the ORF as a function of the phase velocity and gives the expression of the sample variance associated with PTA analyses. Sec.~\ref{sec2} recalls the methodology of the Fisher forecast and how it is applied in our work, and gives the sensitivity needed to distinguish potential phase velocity deviations from GR, with and without considering sample variance. Sec.~\ref{Sec3} draws an estimate of the evolution of the measurement precision of the current PTA collaboration in the next decade and concludes about what deviations could be detected on these time scales.

Throughout the paper, we use the most positive metric signature $(-,+,+,+)$ and natural units such that $c=\hbar=1$.

\section{PTA analysis in modified gravity}
\label{sec1}
The PTA analysis is one of the ways to detect and measure the gravitational wave background present in the interstellar medium. The principle of this analysis is as follows: the gravitational wave background induces slight perturbations in the space-time position of the Earth, which manifest themselves as redshifts in the signals received from pulsars. Computing the correlations between pulsar pairs allows us to detect this common effect and estimate its amplitude. Using an array of pulsars is an effective way to suppress white noise and reduce timing-model uncertainties, making pulsar timing arrays a powerful probe for testing small deviations from general relativity. In this section, we describe the formalism of commonly used PTA analyses~\cite{Anholm_2009} in the context of modified gravity. We work in the frequency domain using the Fourier transform on variables with respect to time.

\subsection{The gravitational waves in modified gravity}
\label{subsec1.A}
To describe the induced perturbations in the pulsar signal, it is useful to decompose the GWB into spherical harmonics, as is often done in CMB polarization analyses. We follow the formalism described in~\cite{Gair_2014, Liang_2023} and assume the GWB to be stationary, isotropic, and uncorrelated, and we approximate the gravitational waves to plane waves.  

Consider the Minkowski spacetime with gravitational waves in the transverse and traceless (TT) gauge, 
\begin{align}
	ds^2 = -dt^2 + d\bm{x}^2 + h_{ij}(t,\bm{x})dx^i dx^j~,
\end{align} 
where $h_{ij}(t,\bm{x})$ denotes the gravitational wave propagating in $\bm x / |\bm x| = \hat\Omega$ direction. It contains two polarization modes and can be expressed in the polarization basis $h_{ij} = h_+ e^+_{ij} +h_\times e^\times _{ij}  $, where the polarization tensors are, 
\begin{equation}
    \begin{aligned}
& e_{i j}^{+}(\hat{\Omega})=\hat{m}_i \hat{m}_j-\hat{n}_i \hat{n}_j\ , \quad e_{i j}^{\times}(\hat{\Omega})=\hat{m}_i \hat{n}_j+\hat{n}_i \hat{m}_j \ ,\\
& \hat{\Omega}=(\sin \theta \cos \phi, \sin \theta \sin \phi, \cos \theta) \ , \\
& \hat{m}=(\sin \phi,-\cos \phi, 0), \hat{n}=(\cos \theta \cos \phi, \cos \theta \sin \phi,-\sin \theta)\ .
\end{aligned}
\end{equation}
Modified gravity theories usually exhibit extra polarization modes and a modified dispersion relation. Since the energy radiated through extra polarization modes is source- and mechanism-dependent, and often suppressed~\cite{deRham:2012fw,Joyce:2014kja}, we focus on the tensor modes ($+,\times$) with a modified dispersion relation in this paper. As the GWB is approximated to be plane waves coming from every direction, the only relevant parameter entering the observable as discussed below is the phase velocity of the gravitational wave~\cite{Liang_2023,Hu:2024wub}. 
 
\subsection{The pulsar signal }
\label{subsec1.B}
We now review the PTA response to the GWB in the framework of modified gravity. 
We consider a single pulsar $P(t_P,\bm{x}_P)$ sending pulses to Earth $E(t,\bm{x}_E)$, for which we have an accurate timing model that predicts a pulse frequency $\nu_0$ at the emission time $t_P$. For these pulses, we observe on Earth a slightly perturbed time-dependent frequency $\nu_E$. The induced redshift can be obtained through the null geodesics in the perturbed spacetime, 
\begin{eqnarray}
\label{reds_time}
    z(t, \hat p)&=&\frac{\nu_0-\nu_E(t)}{\nu_0}= -\frac{1}{2}\int d^2\hat\Omega \,\frac{\hat p^i\hat p^j }{1+\frac{1}{v_p}\hat\Omega\cdot\hat p}\Delta h_{ij} \nonumber\\
    &= & -\frac{1}{2}\int df  d^2\hat\Omega~    e^{2\pi i f t}\left(1-e^{-i 2 \pi f L\left(1+\frac{1}{v_{p }} \hat{\Omega} \cdot \hat{p}\right)}\right) \frac{\hat p^i\hat p^j }{1+\frac{1}{v_p}\hat\Omega\cdot\hat p} h_{ij }\left(f, \frac{1}{v_{p }} \hat{\Omega}\right) \ ,    
\end{eqnarray}
where $\Delta h_{ij}$ denotes the difference of the gravitational wave at the pulsar and Earth, and setting  $\bm{x}_E= 0, \ \bm{x}_P =  L \hat p$ as coordinates for Earth and the pulsar. The $1/v_p$ in the denominator of Eq.~\eqref{reds_time} represents the correction due to the modified dispersion relation~\cite{Liang_2023}.   
The relevant quantity to study in PTA analyzes is the residual of arrival time, defined as the integral of the redshift over the observation time, 
\begin{equation}
\label{defres}
    R(t,\hat p)=\int_0^T dt'z(t',\hat p)\ .
\end{equation}  
In real PTA analysis, the residuals of arrival times are analyzed in discrete frequency bins. For now, we set aside the discussion of discrete Fourier transformations and present the continuous Fourier transformation,  
\begin{equation}
\begin{aligned}
\label{res_int}
R(f,\hat p ) &\equiv  \int df e^{2\pi i ft} R(t,\hat p) \\ &=   \frac{1}{2\pi i f} \int d^2\hat{\Omega} \, \frac{\hat{p}^i \hat{p}^j}{2\left(1 + \frac{1}{v_p} \hat{\Omega} \cdot \hat{p} \right)} \times\left( 1 - e^{-2\pi i f L \left( 1 + \frac{1}{v_p} \hat{\Omega} \cdot \hat{p} \right)} \right)
h_{ij }\left(f, \frac{1}{v_{p }} \hat{\Omega}\right) \ , 
\end{aligned}
\end{equation} 
where we assume the observation time $T\to \infty$. The finite observation and resulting discrete Fourier transform will be revised later when discussing the sample variance effect\footnote{The sample variance sometimes is also referred to as cosmic variance, by analogy with cosmic microwave background analysis in the literature. }. 
 
\subsection{Correlation signal} 
Since the stochastic GWB is a common process that affects all pulsar signals, our goal is to compute the correlation between the residuals of a pulsar pair. Assuming an isotropic GWB, we can separate the correlation between two pulsars into the power spectrum and overlap reduction function (ORF),  
\begin{eqnarray}
\label{eq,commonspec}
 \Phi_{12}(f,f',\xi)& \equiv& \left\langle R^*(f, \hat{p}_1)\, R(f', \hat{p}_2) \right\rangle = \delta(f-f')\,H(f)\,\Gamma(f,\xi)\ ,
\end{eqnarray}
where $\xi = \cos^{-1}(\hat p_1 \cdot \hat p_2)$ is the angular separation between pulsar pairs, and $H(f)$ is the power spectrum, 
\begin{equation}
    \left\langle h^{A*}(f, \hat{\Omega}) \, h^{A'}(f', \hat{\Omega}') \right\rangle 
= \frac{1}{2}\delta^2(\hat{\Omega}, \hat{\Omega}') \, \delta_{A A'} \, \delta(f - f') \, H(f) \ .
\end{equation}
Here $A,A'$ run over the polarizations $\{+,\times\}$.  
$\Gamma(f,\xi)$ in Eq.~\eqref{eq,commonspec} is the overlap reduction function (ORF) that encodes the information of angular separation. 
In general relativity, after taking the limit $fL \gg 1$, we are safe to neglect the fast-oscillating pulsar term in Eq.~\eqref{reds_time} when computing the two-point correlation function due to decoherence, and the ORF turns to the famous Hellings-Downs curve~\cite{Hellings:1983fr},  
\begin{equation}
\label{def_HD}
    \text{HD}(\xi )= \frac{3+\cos{\xi }}{8} +  \frac{3(1-\cos{\xi })}{4} \log \frac{1-\cos{\xi }}{2}  \ .  
\end{equation}  
In modified gravity, the ORF is altered by the presence of extra polarization modes and modified dispersion relations. As discussed in the previous section, we will only focus on the modified phase velocity of the tensor modes. The ORF can be decomposed into a Legendre polynomial basis, with the coefficients taking the form~\cite{Liang_2023}
\begin{eqnarray}
\label{def_ORF}
  \Gamma(f,\xi)= \mathcal{C} \sum_{\ell = 2}  2(2\ell+1)\frac{(\ell-2)!} {(\ell+2)!}\left|c_\ell(f)\right|^2 \,P_\ell(\cos \xi )\equiv  \sum_{\ell = 2}   \frac{2\ell+1}{4\pi} C_\ell (f)\,P_\ell(\cos \xi ) \ , 
\end{eqnarray}
where $P_\ell$ represents the Legendre polynomial of degree $\ell$, and $\mathcal{C}$ is a normalization constant that can be absorbed into the power spectrum. We will specify its choice at the end of this section.   
\begin{eqnarray}
     C_\ell (f) \equiv 8\pi \mathcal{C} \frac{(\ell-2)!} {(\ell+2)!}\left|c_\ell(f)\right|^2  \ ,
\end{eqnarray}
is defined to match the coefficients in literature~\cite{Roebber:2016jzl,Bernardo:2022xzl}. The frequency dependence in the ORF arises from two factors: the pulsar term, which can be neglected in the long-distance limit, and the implicit frequency dependence of the phase velocity. $c_\ell(f)$ denotes the modification arising from the phase velocity,
\begin{equation}
    c_\ell(f)=\int_{-1}^{1} dx \, \frac{(1 - x^2)^2}{1 + \frac{1}{v_p} x}\left[ 1 - e^{-i 2 \pi f L \left(1 + \frac{1}{v_p} x \right)} \right]\frac{d^2}{dx^2} P_\ell(x)\ ,
\end{equation}
recovering the general relativity result of $4 (-1)^\ell$ in the long-distance limit $fL \gg 1$ and $v_p \to 1$.  In the above expression, the integral on $[-1,1]$ comes from the choice of a coordinate frame where $\hat p$ is aligned with the $z-$axis, so that $\hat\Omega\cdot\hat p=\cos\theta\equiv x$. For $v_p \leq 1$, the exponential term cannot be neglected even in the long-distance limit, as it is necessary to cancel the apparent pole in the denominator. 
This expression reads~\cite{Liang:2024mex}, 
\begin{equation}
\label{approx_cl}
c_\ell = 
\begin{cases}
\begin{aligned}
-2v_p(1+\ell)\Big(&[(2+\ell)v_p^2 - \ell]\, \tilde Q_\ell(-v_p) + 2v_p\, \tilde Q_{\ell+1}(-v_p)\Big) + \mathcal{O}\left(\frac{v_p}{fL}\right)\ ;
\end{aligned}
& v_p > 1\ , \\[2ex]
4(-1)^\ell + \mathcal{O}\left(\frac{v_p}{fL}\right)\ ;
& v_p = 1\ , \\[2ex]
\begin{aligned}
-2v_p(1+\ell)\Big(&[(2+\ell)v_p^2 - \ell]\, Q_\ell(-v_p)  + 2v_p\, Q_{\ell+1}(-v_p)\Big) \\
+ \pi i v_p(1+\ell)\Big(&[(2+\ell)v_p^2 - \ell]\, P_\ell(-v_p) + 2v_p\, P_{\ell+1}(-v_p)\Big) + \mathcal{O}\left(\frac{v_p}{fL}\right)\ ;\\
\end{aligned}
& v_p < 1 \ . 
\end{cases}
\end{equation}
 
In this expression, $\tilde Q_\ell(x)$ and $Q_\ell(x)$ are Legendre functions of the second kind of degree $\ell$, associated with solutions for $x\in \mathbb{R}\setminus[-1,1]$ and $x\in(-1,1)$, respectively. At large $\ell$, for $v_p > 1$, $|c_\ell|^2$ contains an exponential decay; therefore, the contribution from higher multipole moments decays rapidly. To compute the series numerically in this case, we cut the computation to a sufficiently large $\ell$ ($\approx 50$) depending on $v_p$. For $v_p < 1$, utilizing the approximation~\cite{liang2024testinggravityfrequencydependentoverlap},  
\begin{equation}
    \lim_{\ell\to\infty} 2(2\ell+1)\frac{(\ell-2)!} {(\ell+2)!}\left|c_\ell(f)\right|^2   = 8\pi v_p^2(1-v_p^2)^{3/2}\ ,
\end{equation}
and the Legendre polynomial identity
\begin{equation}
\label{P_l_sum_id}
    \sum_{\ell=0}^\infty P_\ell(\cos \xi)=\frac{1}{\sqrt{2-2\cos\xi}}\ ,
\end{equation}
we can express the ORF for the subluminal case as
\begin{eqnarray}
    \Gamma(\xi,v_p<1) &=&  \bigg(\frac{8\pi v_p^2(1-v_p^2)^{3/2}}{\sqrt{2-2\cos\xi}} + \sum_{\ell=2} \Big(2(2\ell+1)\frac{(\ell-2)!} {(\ell+2)!}\left|c_\ell(f)\right|^2- 8\pi v_p^2(1-v_p^2)^{3/2}\Big) P_\ell(\cos\xi)  \nonumber\\
    && ~~  - 8\pi v_p^2(1-v_p^2)^{3/2}(P_0(\cos\xi) + P_1(\cos\xi) ) \bigg) \mathcal{C}\ .
\end{eqnarray}

The naive divergence as $\xi \to 0$ is unphysical and arises from the infinite-distance assumption used in computing Eq.~\eqref{approx_cl}. The finite-distance correction is a non-perturbative effect~\cite{Domenech:2024pow}, which introduces a physical cut-off in the multipole summation at $\ell \sim f L/v_p$, beyond which the multipole coefficients decay exponentially, rendering the ORF finite~\cite{liang2024testinggravityfrequencydependentoverlap}. Although the physical interpretation is unaffected, since a real PTA system always has a minimal angular separation, this divergence complicates the choice of the normalization factor $\mathcal{C}$~\cite{Anholm_2009,Agazie_2023,cordes2025overlapreductionfunctionpulsar}, which is conventionally set so that $\Gamma(\xi = 0) = 1/2$. Throughout this paper, we normalize the ORF at $\xi = 180^\circ$ to $1/4$ to match the GR prediction. We will show in Sec.~\ref{sec2} and in Appendix~\ref{apx_ORF} that this is a justified normalization choice.   

Having established the theoretical ORF in the long-distance limit as a function of phase velocity, we now examine the intrinsic noise before analyzing the sensitivity. 
\subsection{Sample variance effect}
\label{1.D} 
In the previous section, we studied the modified overlap reduction function arising from the modified dispersion effect. Our goal is to forecast the uncertainty in measuring deviations from GR in next-decade PTA surveys. Therefore, understanding the systematics involved in measuring and reconstructing the ORF is essential for this task. 

Various effects, including the precision of measuring the residual of arrival time of pulses, can lead to a fluctuation in the reconstruction of the ORF. We discuss two variance effects here. One is the pulsar variance, which arises because we have a finite number of pulsars that are not isotropically distributed in the sky. This effect can be reduced by finding more milli-second pulsars and approaching the higher multipole moment in harmonic analysis~\cite{Nay:2023pwu}. 

\begin{figure}   
\centering
\includegraphics[scale=0.45 ]{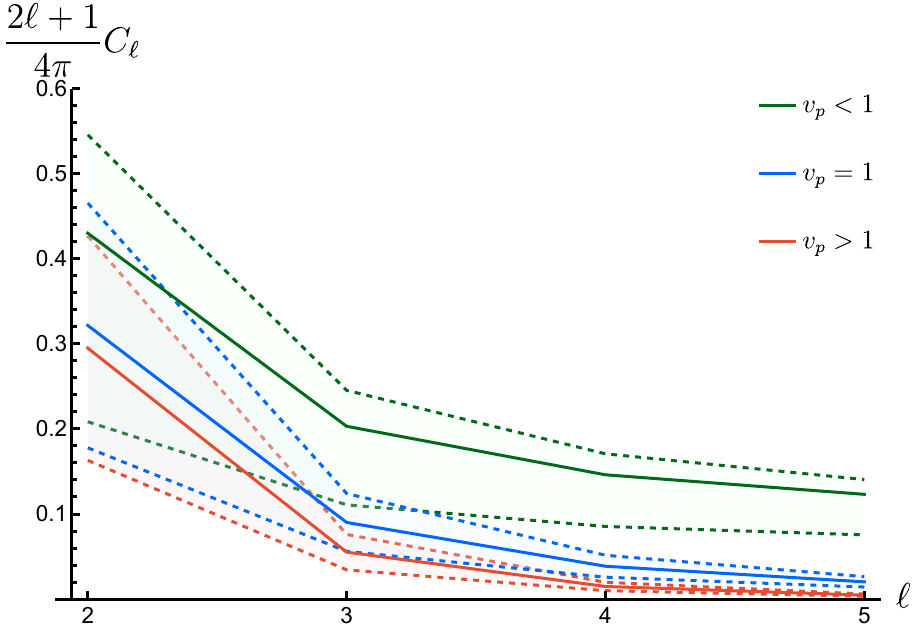}~~ ~~~
\includegraphics[scale=0.4 ]{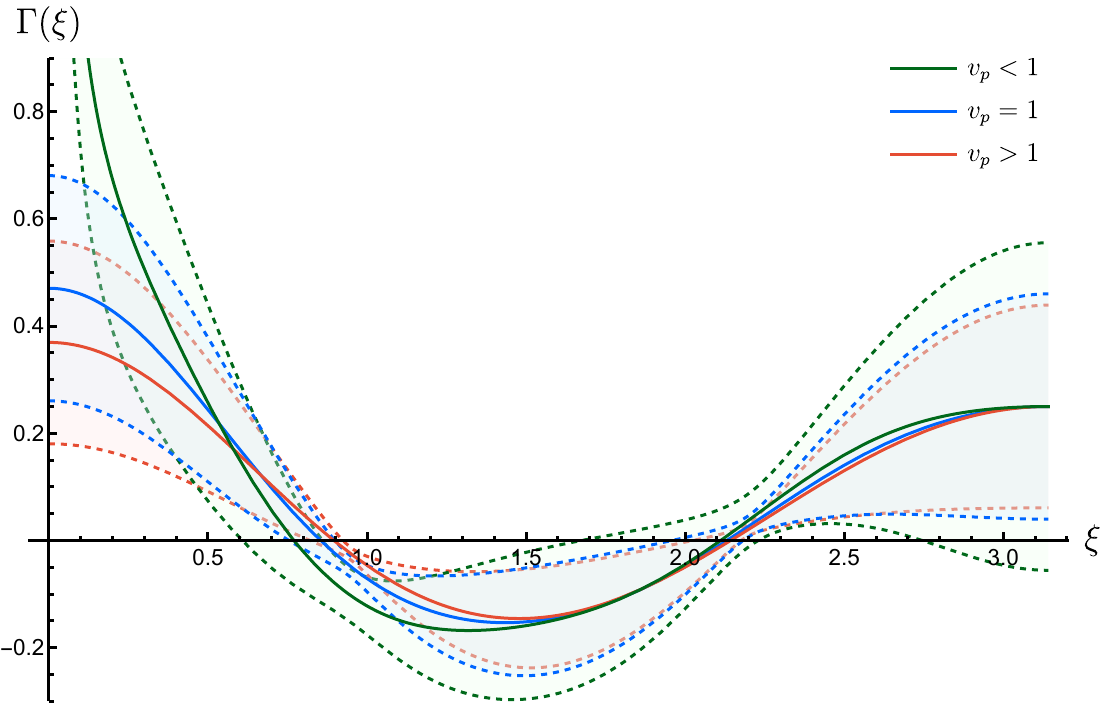}
    \caption{The sample variance in modified gravity vs. GR. The right panels show the overlap reduction function $\Gamma(\xi)$, and the left panels show the coefficients of the Legendre polynomials. The blue lines and shaded areas represent the ORF for GR and its variance, the red lines correspond to the massive-gravity–type modified dispersion with $v_p > 1$, and the green lines correspond to the case with $v_p < 1$.}
    \label{Fig_sample_variance}
\end{figure}

The second effect is the sample variance, also known as cosmic variance in the literature, aiming to draw an analogy with CMB analyzes~\cite{Allen:2022dzg,Roebber:2016jzl,Bernardo:2022xzl}. This is harder to remove since it arises from the intrinsic noise from the distribution of sources. Due to the fundamental stochasticity of the underlying GWB, the reconstruction of the HD curve in GR exhibits fluctuations even when assuming an infinite number of pulsars. These fluctuations reflect the fact that, because of the finite observation time and the resulting frequency bins, interference between different sources can fall into the same bin. 
As discussed in~\cite{Allen:2024uqs}, the procedure of obtaining the frequency bin in real data analysis involved a $\text{sinc} (\pi T (f-f_k)) = \sin (\pi T (f-f_k)) / \pi T (f-f_k)) \neq \delta (f- f_k)/T $ function in the $k-$th frequency bin, and therefore, introduce the interference between different sources. These interference introduces variance in the ORF~\cite{Bernardo:2022xzl,Allen:2022dzg}, 
\begin{eqnarray}
    \sigma_{\text{SV}}^2= \braket{\Gamma(f,\xi)^2} -   \braket{\Gamma(f,\xi)}^2 = \sum_{\ell=2} \frac{2 \ell+1}{8 \pi^2} C_\ell^2 P_\ell^2(\cos \xi)\ ,
\end{eqnarray}
where the bracket denotes the sum over all spherical harmonics, and this variance corresponds to the upper panel of Fig.~\eqref{Fig_sample_variance}, where we plot $\Gamma(f,\xi) \pm \sigma_{\text{SV}}$. 
In the angular power spectrum, it shows as~\cite{Roebber:2016jzl}\footnote{Notice the $\sqrt{2}$ difference compared to~\cite{Bernardo:2022xzl}. }, 
\begin{equation}
\label{eq,DeltaCl}
    \Delta C_\ell = \frac{C_\ell}{\sqrt{2\ell+1 }}\ . 
\end{equation} 

We show in Fig.~\eqref{Fig_sample_variance} the sample variance effect for GR (blue), massive gravity with $v_p = 1.1$ (red), and the subluminal case $v_p = 0.9$ (green). One can see that the difference between GR and massive gravity mostly falls within the variance band, making their separation especially difficult. The physical reason is straightforward: when $v_p > 1$, the coefficients of higher multipole moments are suppressed, and the sample variance is dominated by the quadrupole moment. This is clearer in harmonic analysis, where we explicitly plot the coefficients for the first five multipole moments in Fig.~\eqref{Fig_sample_variance} as suggested in~\cite{Nay:2023pwu}. Starting from the octopole moment, the difference becomes more apparent.  
 
Notice that the above expression is primarily applicable when analyzing a single frequency bin, as in current analyses of the data by NANOGrav. As noted in~\cite{Nay:2023pwu,Allen:2024uqs}, the sample variance can be reduced by considering cross-frequency analyses. However, as we will show later, in the regime of interest the sample variance is not the dominant source of noise. Moreover, the optimal reconstruction of the overlap reduction function for modified gravity theories remains underexplored. Therefore, we adopt the simpler form, Eq.~\eqref{eq,DeltaCl}, for the remainder of this paper.

\section{Fisher forecast methodology}
\label{sec2}

We have shown in Sec.~\ref{sec1} that the deviation of the overlap reduction function from the Hellings-Downs curve can serve as a probe to identify modified gravity. Since we know how this observable is modified under a deviation in the phase velocity, it is worth investigating how feasible it is to detect such deviations from GR in future decades-long surveys. To assess the uncertainty $\sigma_v$ on the measurement of the gravitational wave phase velocity in the next years, we use the Fisher formalism. This method aims to determine how much an observable is sensitive to the model input parameters. In our case, our observable is the overlap reduction function, and our only input parameter is the phase velocity. In this section, we apply this methodology to relate the forecast uncertainty $\sigma_v$ to the standard deviations of pulsar pair correlation measurements derived from the optimal detection statistic~\cite{Anholm_2009,Agazie_2023}.

\subsection{The Fisher information}
\label{sec2.A}

In the context of modified gravity, assuming the plane wave nature of the GWB and focusing on the tensor modes, we have shown in Sec.~\ref{sec1} that the only parameter that enters the overlap reduction function is the phase velocity, $v_p$. Following this result, we apply the Fisher formalism to the observable $\Gamma (\xi,v_p)$ with respect to the parameter $v_p$. Assuming an underlying real value of the phase velocity $v_p^{real}$, we compute Fisher information to obtain an estimation of the uncertainty on $v_p^{real}$ in the measurement. This uncertainty is the best precision we can have for this parameter. 

To compute the uncertainty $\sigma_v$, we choose a fiducial velocity $v_p^{fid}$ that we think is the underlying real value $v_p^{real}$. Given a measurement of correlations among pulsar pairs with angular separation $\xi_{ij}$ and measurement uncertainty $\sigma_{ij}$, we define Fisher information as
\begin{equation}
\label{fisher_def_theoric}
    \mathcal{F}(v_p^{fid})=\mathbf{E}\left(\left(\frac{\partial}{\partial v_p}\log \mathcal{L}\,(\Gamma^{obs};v_p)\right)^2\Big |\,v_p^{fid}\right) ,
\end{equation}
where 
\begin{eqnarray}
\label{loglike}
    &\log\mathcal{L}\,(\Gamma^{obs};v_p)=-\frac{1}{2}\left(\Gamma^{obs}-\Gamma(v_p)\right)^T\Sigma^{-1}\left(\Gamma^{obs}-\Gamma(v_p)\right)-\frac{1}{2}\log \left(2\pi\det\Sigma\right)\ ,
\end{eqnarray}
is the log-likelihood of the Gaussian random multivariate $\Gamma^{obs}$ (the observed correlations) with respect to the parameter $v_p$, and $\mathbf{E}\left(\mathbf{X}\,\big |\,v_p^{fid}\right)$ is the expectation value of $\mathbf{X}$ assuming that $v_p^{fid}$ is the true parameter.  Here $\left(\Gamma^{obs}-\Gamma(v_p)\right)_{ij}\equiv\Gamma^{obs}_{ij}-\Gamma(\xi_{ij},v_p)$, where $i,j$ run over the pairs of pulsars $\braket{i,j}$ and $\Sigma$ is the covariance matrix of the measured correlations $\Gamma^{obs}_{ij}$. We will first assume a spatially uncorrelated distribution for the correlations $\Gamma^{obs}$, so that $\Sigma$ is diagonal\footnote{ Since the noise of pulsars is spatially independent, it is a well-justified assumption that the covariance matrix is diagonal.}, and $\Sigma=\Sigma(v_p)$ in general. 
Under those assumptions, it is easy to derive the well-known expression of the Fisher information for a parameter-dependent covariance matrix
\begin{equation}
    \label{def_fisher_theory}
    \mathcal{F}(v_p^{fid})=\left(\frac{\partial\Gamma}{\partial v_p}\bigg|_{v_{p}=v_{p}^{fid}}\right)^T\Sigma^{-1}\left(\frac{\partial\Gamma}{\partial v_p}\bigg|_{v_{p}=v_{p}^{fid}}\right)+\frac{1}{2}\text{Tr}\left(\Sigma^{-1}\frac{\partial \Sigma}{\partial v_p}\bigg|_{v_{p}=v_{p}^{fid}}\,\Sigma^{-1}\frac{\partial \Sigma}{\partial v_p}\bigg|_{v_{p}=v_{p}^{fid}}\right)\ .
\end{equation}
Fisher information describes how sensitive the ORF is to the parameter $v_p$, at the fiducial value $v_p^{fid}$, relatively to the measurement uncertainties contained in $\Sigma$. In our case, the covariance matrix includes both the uncertainty $\sigma_{ij}$ of the measurement of the correlation between the pulsars $i$ and $j$ and the sample variance $\sigma_{SV}$ derived in \ref{1.D}. The measurement uncertainty $\sigma_{ij}$ is the standard deviation of the optimal detection statistic for pair $\langle i,j\rangle$~\cite{Anholm_2009}. A simplified derivation of $\sigma_{ij}$ is presented in~\cite{Anholm_2009} using the optimal statistic formalism. The covariance matrix $\Sigma$ is a diagonal matrix that reads
\begin{equation}
\label{eq,covariancematrix}
\Sigma=\text{diag}\left(\sigma_{ij}^2+\sigma_{SV}^2(\xi_{ij},v_p)\right)\ .
\end{equation}
The Fisher information gives a lower bound on the uncertainty of the measurement of the true parameter $v_p^{true}$. We denote this uncertainty $\sigma_v^{true}$. If $v_p^{fid}$ is close to the true parameter, we have
\begin{equation}
\label{sigma_fisher}
    \sigma_v^{true}\geq \sigma_v(v_p^{fid})\equiv\frac{1}{\sqrt{\mathcal{F}(v_p^{fid})}}\ .
\end{equation}
In the case of an optimistic forecast, we take $\sigma_v^{true}=\sigma_v$, as is usually done in a Fisher forecast. We show in Appendix~\ref{apx_Fisher} that this is, in fact, a very good approximation. The reason is that we use almost systematically the maximum likelihood estimator of the phase velocity to measure the true parameter in the frequentist approach. We make use of the $\chi^2$ statistic defined as $\chi^2\equiv-2\log \mathcal{L}$ to relate the Fisher information to the curvature of the log-likelihood. If $v_p^{fid}$ is close to the true parameter, the first derivative of the $\chi^2$ vanishes and we can approximate it by a parabola. The Fisher information is related to the curvature of the $\chi^2$ around that point and gives an estimate of the uncertainty $\sigma_v^{true}$ of the measurement of $v_p^{true}$ following Eq.~\eqref{sigma_fisher}.
 
Since our goal is in the first place to provide a null test to GR rather than measuring precisely $v_p^{true}$, we quantify the result in terms of the $\sigma$-distance. It is defined as
\begin{equation}
\label{def_sigma_dist}
d_\sigma(v_p^{fid})\equiv\frac{|v_p^{fid}-1|}{\sigma_v(v_p^{fid})}\ .
\end{equation}
This distance tells us at which confidence level in number of $\sigma$ we are able to exclude GR or not, assuming that the true parameter is close to $v_p^{fid}$. As the numerator of the fraction in Eq.~\eqref{def_sigma_dist} is known for a given $v_p^{fid}$, we will focus on the phase velocity dependence of $\sigma_v$.
In a first analysis, we neglect the sample variance compared to the measurement uncertainty $\sigma_{ij}$. The covariance matrix becomes phase velocity-independent and the Fisher information reduces to 
\begin{equation}
\label{fisher_def}    
\mathcal{F}(v_p^{fid})=\sum_{\langle i,j\rangle}\frac{1}{\sigma_{ij}^2}\left[\frac{\partial\Gamma(\xi_{ij},v_p)}{\partial v_p}\bigg|_{v_{p}=v_{p}^{fid}}\right]^2\ .
\end{equation}
Note that the normalization of the overlap reduction function may affect the Fisher information and $\sigma_v$. However, as shown in Fig.~\eqref{Norm_deviat}, for the phase velocities of interest, the difference arising from the choice of normalization factor is negligible. Further details on how the normalization modifies the forecast are provided in Appendix~\ref{apx_ORF}. We can also choose a different statistic to theoretically avoid this normalization dependence. For example, in the CPTA collaboration~\cite{Xu_2023}, a significance function $\mathcal{S}$, derived from the Pearson correlation, is used to quantify the statistical significance. However, the uncertainties $\sigma_{ij}$ do not appear explicitly in the associated observable information, making the forecast complicated. We save further discussion in Appendix~\ref{apx_S_fun}. We therefore stick to the Fisher formalism in the rest of the analysis. 

The forecast uncertainty $\sigma_v$ depends on three parameters. First, the number of pulsars determines the number of terms in the sum, so increasing the number of pulsars reduces the uncertainty $\sigma_v$. The other two parameters are the fiducial phase velocity $v_p^{fid}$, which affects the sensitivity of the observable, and the measurement uncertainties $\sigma_{ij}$ for each pair of pulsars. To separate the effect from fiducial velocity and the precision and simplify our forecast, we assume a universal uncertainty $\bar\sigma$ for all pulsar pairs. It can be defined through the inverse quadratic mean 
\begin{eqnarray}
\label{eq,defbarsigma}
\bar\sigma \equiv \sqrt{ N_{pairs}}\bigg/\sqrt{\sum_{< ij> } \frac{1}{\sigma_{ij}^2} }\ ,
\end{eqnarray}
and captures the ability of the radio telescope.  
The uncertainty of the phase velocity prediction is proportional to this value, $\sigma_v \propto \bar\sigma$. The more precisely we can measure each individual pulsar pair, the more accurately we can distinguish the phase velocity of gravitational waves from the speed of light. 

To estimate $\bar \sigma$ for a certain PTA system, we need information on the spatial distribution of pulsars in the array and the precision of their timing measurements carried out by $\sigma_{ij}$. We have derived through Eq.~\eqref{fisher_def} and \eqref{sigma_fisher} the explicit relation between the (minimal) uncertainty $\sigma_v$ on the measurement of a given fiducial velocity and the uncertainties of the measurement of pulsar pair correlations $\sigma_{ij}$. In Eq.~\eqref{fisher_def}, the forecast uncertainty $\sigma_v$ is evidently sensitive to both the spherical distribution of the pulsars (through its dependence on $\xi_{ij}$) and the relation between $\sigma_{ij}$ and $\xi_{ij}$. 
\begin{figure}[h]
    \centering
\includegraphics[width=\linewidth]{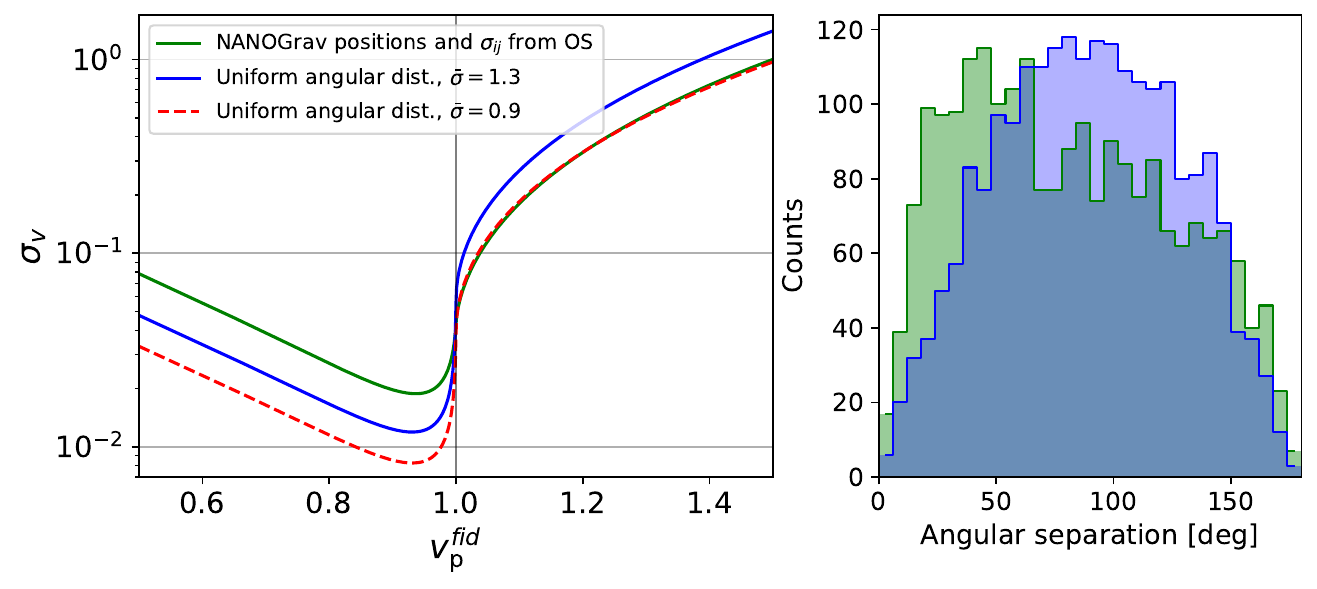}
    \caption{\textit{Left panel:} Forecast $\sigma_v$ (log-scaled) from the Fisher formalism as a function of $v_p^{fid}$, using the angular separations $\xi_{ij}$ and uncertainties $\sigma_{ij}$ from the NANOGrav's $15$ yrs dataset maximum-likelihood noise parameters optimal statistic~\cite{Agazie_2023,the_nanograv_collaboration_2025_16051178} (green curve), and a random angular distribution following Eq.~\eqref{pdf_xi} and constant $\bar\sigma$ (blue and red dashed curve). We use the same number of pulsars ($N_{psr}=67$) for all curves and the same positions for the blue and dashed red curves. For the blue curve, we choose $\bar\sigma=\sqrt{\frac{N_{pairs}}{\sum_{\langle i,j\rangle}1/\sigma_{ij}^2}}\approx1.3$ where $\sigma_{ij}$ are the uncertainties from the NANOGrav's optimal statistic. For the red dashed curve, we choose $\bar\sigma=0.9$ such that the red and green curves overlap for $v_p>1$. \textit{Right panel:} Distribution of the angular separations associated with the blue (uniformly distributed) and green curves (NANOGrav's data) in the left-hand side plot.}
    \label{Comp_NG_rand} 
\end{figure}
To keep our forecast as general as possible while simplifying the calculations, we assume a constant $\sigma_{ij}\equiv\bar\sigma$ with respect to $\xi_{ij}$ and a random uniform spherical distribution for the pulsars as in~\cite{Allen_2023}. In this framework, the distribution of angular separations follows the probability density function
\begin{equation}
\label{pdf_xi}
    p(\xi_{ij})=\frac{1}{2}\sin(\xi_{ij})\ ,
\end{equation} 
for $\xi_{ij}\in[0^\circ,180^\circ]$ as shown in the blue distribution in the right panel of Fig.~\eqref{Comp_NG_rand}; however, the real system may deviate from a uniform distribution, as illustrated by the green region. In the left panel of Fig.~\eqref{Comp_NG_rand}, we compare the forecast uncertainty $\sigma_v$ in two scenarios: constant values of $\bar\sigma$ with uniformly distributed pulsars (blue and red curves), and the real pulsar distribution from NANOGrav’s 15-year dataset (green curve)~\cite{Agazie_2023,the_nanograv_collaboration_2025_16051178}. We can see that the forecast uncertainty depends on the pulsar distribution, naturally, because the angular dependence in the ORF causes each angle to encode different information.

Here, $\bar \sigma = 1.3 $ for the blue curve is obtained from the inverse quadratic mean of the uncertainties of the NANOGrav optimal statistic with maximum likelihood noise parameters as defined in Eq.~\eqref{eq,defbarsigma}.  We see that the uncertainty is larger than that of the real distribution in the $v_p > 1$ regime, and smaller than the real distribution in the $v_p < 1$ regime. If we want to focus on the case with phase velocity greater than 1, we can choose a smaller $\bar\sigma = 0.9$ to better represent the current uncertainty level in this regime, as shown by the red curve. 
Since the actual distribution of future pulsars over the sky is unknown, we will maintain the assumption of uniformly distributed pulsars and use constant values of $\bar\sigma$ in the following forecast analysis. 
 
Another key factor for the forecast sensitivity is the number of observed pulsars, which we next use to scale our predictions. The uncertainty $\sigma_v$ decreases linearly with $N_{psr}$ for large number of pulsars, 
\begin{equation}
\label{F_prop_Npair}
\sigma_v \propto \frac{\bar\sigma}{\sqrt{N_{pairs}}}\approx\frac{\bar\sigma}{N_{psr}}\ ,
\end{equation}
where $N_{pairs}=N_{psr}(N_{psr}-1)/2$ is the number of pulsar pairs in the system.

To separate this scaling from the phase velocity, we introduce the reduced uncertainty
\begin{equation}
\label{red_sigma}
    \tilde\sigma\equiv \bar\sigma\,\frac{67}{N_{psr}}\ ,
\end{equation}
with respect to the Fisher information at a fixed pulsar number of $67$. We define $\mathcal{F}_{67}$ with respect to the current data to quantify the effect from the phase velocity, under the assumption that the precision of each pulsar pair is $\sigma_{ij} = 1$, 
\begin{eqnarray}
   \mathcal{F}_{67}(v_p^{fid})\equiv\sum_{\substack{\braket{i,j}\\N_{psr}=67}}\left[\frac{\partial\Gamma(\xi_{ij},v_p)}{\partial v_p}\bigg|_{v_{p}=v_{p}^{fid}}\right]^2\ .
\end{eqnarray}
Therefore, we can express the forecast uncertainty $\sigma_v$ as 
\begin{equation}
\label{eq,reduceuncertainty}
    \sigma_v=\frac{\tilde\sigma}{\sqrt{\mathcal{F}_{67}(v_p^{fid})}}\ . 
\end{equation} 
This definition absorbs all adjustable physical parameters into $\tilde\sigma$, leaving the dependence on the fiducial velocity $v_p^{fid}$ only in $\mathcal{F}_{67}$. A direct proportionality relates $\sigma_v$ and $\tilde\sigma$, so the reduced uncertainty $\tilde\sigma$ becomes the primary quantity for comparing our forecast with the properties of current surveys, such as total observation time or timing precision. 

The final step before applying this methodology to estimate the required $\tilde\sigma$ is to validate the Fisher approximation. 
In Sec.~\ref{apx_backtest}, we perform a mock-data analysis to assess the validity of the Gaussian approximation underlying the Fisher forecast. We find that the Fisher information provides a reasonable estimate of $\sigma_v$ for the recovery of the phase velocity, while for larger values of $v_p^{fid}$, non-Gaussian features of the likelihood become important and affect the comparison with the GR prediction. We then proceed in the next section to carry out the forecast as outlined above.

\subsection{Forecast on parameter sensitivity}
\label{subsec3.B}
In this section, we study the reduced uncertainty $\tilde{\sigma}$ required to identify deviations from GR. For the $n$-$\sigma$ confidence level that we desire, we want to find the largest uncertainty such that $n\leq d_\sigma(v_p^{\mathrm{fid}})$, where $d_\sigma(v_p^{\mathrm{fid}})$ is the $\sigma$-distance defined in Eq.~\eqref{def_sigma_dist}. This, together with Eq.~\eqref{eq,reduceuncertainty}, gives 
\begin{equation}
\label{nsigma_deviation}
    \tilde{\sigma}_{max}=\frac{1}{n}\sqrt {\mathcal{F}_{67}\left(v_p^{fid}\right)}\,\left|v_p^{fid}-1\right|\ .
\end{equation}
We use this equation to determine the requirements for detecting a potential deviation from GR at a given confidence level. 

 Table.~\eqref{tab_sigma_req} shows the required $\tilde\sigma_{max}$ for different fiducial values. As expected, we need higher precision (lower uncertainty) for the $v_p>1$ case. We highlight in green the parameters for which the necessary uncertainty exceeds $1.3$. It may appear that phase velocities below $0.95$ could already be excluded at the $3\sigma$ confidence level given the current precision; however, this is not the case, as the table presents only an optimistic forecast and the real $\sigma_v$ is undervalued (see left panel of Fig.~\eqref{Comp_NG_rand}). However, it illustrates the asymmetry between the $v_p < 1$ and $v_p > 1$ cases, indicating that the former is more likely to be excluded as the data improve.

\begin{table}[h]
\centering
\renewcommand{\arraystretch}{1}
\begin{tabular}{|c||c|c|c||c|c|c|c|c|}
\hline
 Deviation from GR& $-10\%$ & $-5\%$ & $-1\%$ & $+1\%$ & $+5\%$ & $+10\%$ & $+20\%$ & $+50\%$ \\
\hline
$1\sigma$ 
 & \cellcolor{green!25}10.524 & \cellcolor{green!25}5.333 & 0.677 & 0.138 & 0.376 & 0.486 & 0.541 & 0.462 \\
\hline
$2\sigma$ 
 & \cellcolor{green!25}5.262 & \cellcolor{green!25}2.666 & 0.339 & 0.069 & 0.188 & 0.243 & 0.270 & 0.231 \\
\hline
$3\sigma$ 
 & \cellcolor{green!25}3.508 & \cellcolor{green!25}1.778 & 0.226 & 0.046 & 0.125 & 0.162 & 0.180 & 0.154 \\
\hline
$5\sigma$ 
 & \cellcolor{green!25}2.105 & 1.067 & 0.135 & 0.028 & 0.075 & 0.097 & 0.108 & 0.092 \\
\hline
\end{tabular}
\caption{Maximum $\tilde\sigma$ values required to distinguish different phase velocity deviations from GR at $1,2,3$ and $5\sigma$ confidence levels. The shown percentages correspond to a phase velocity deviation. A $+10\%$ deviation corresponds to $v_p^{fid}=1.1$ and a $-10\%$ to $v_p^{fid}=0.9$. Values greater than $\tilde\sigma=1.3$ computed from NANOGrav 15yr dataset~\cite{Agazie_2023} are highlighted in green.}
\label{tab_sigma_req}
\end{table}
  
It is important to note that this Fisher forecast may vary depending on the PTA collaboration considered, due to the angular repartition and the specific precision of each pulsar pair in the array. For example, in the case of NANOGrav, the ORF appears less sensitive to $v_p < 1$ than in our forecast, implying that deviations from GR in this regime may be harder to detect. In contrast, the sensitivity increases for $v_p > 1$, making such deviations easier to distinguish.
 
\begin{figure}[h]
    \centering
    \includegraphics[width=.9\linewidth]{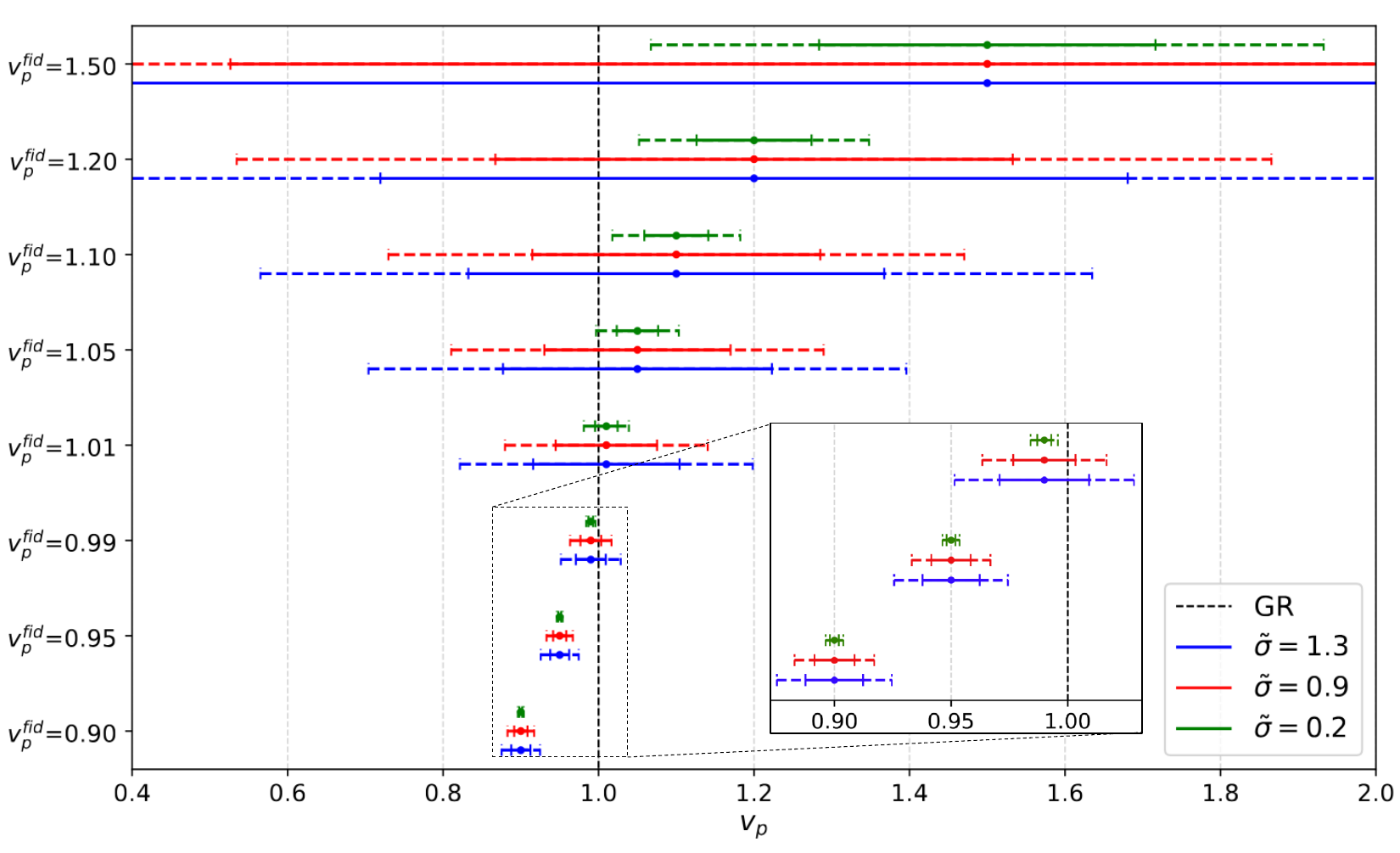}
    \caption{Whisker plot of the forecast confidence levels for different $v_p^{fid}$ and uncertainties $\tilde\sigma$. The $1\sigma$ (solid lines) and $2\sigma$ (dashed lines) are shown for comparison to the GR case ($v_p=1$, dashed black line). For large $v_p^{fid}$ and $\tilde\sigma$, the errors bars exceed the window width. Positive deviations ($v_p^{fid}>1.1$) only start to be distinguishable at $2\sigma$ level for $\tilde\sigma=0.2$. }
    \label{whisk_fcst}
\end{figure}
To better understand how distinguishable a given $v_p^{fid}$ is from GR, we show in Fig.~\eqref{whisk_fcst} a whisker plot for different deviations. For each deviation, we show the $1\sigma$ (solid) and $2\sigma$ (dashed) confidence intervals around the fiducial mean value $v_p^{fid}$, so that $95\%$ of the data lie within the interval $[v_p^{fid}-2\sigma,~ v_p^{fid}+2\sigma]$. If the interval does not overlap with $v_p=1$, then we can distinguish this deviation from GR at more than the $2\sigma$ level. The confidence intervals are shown for three different precisions, $\tilde\sigma=1.3$ (in blue), $\tilde\sigma=0.9$ (in red), and $\tilde\sigma=0.2$ (in green). As we can see, the current precision can already probe the deviation between $v_p^{fid}=0.9$ or $0.95$ and GR at the $2\sigma$ level, whereas if the underlying dispersion relation has a phase velocity greater than $1$, it will require the precision to improve by a factor of 4, $\tilde\sigma = 0.2$.

The symmetric error bar is a feature of the Fisher forecast, where the underlying data sets are assumed to be Gaussian. For the current uncertainties, $\tilde{\sigma}=1.3$ and $0.9$, shown by the blue and red lines respectively, the superficial crossover to superluminal propagation in the $v_p>1$ case can be confusing, as the error bars for $v_p=0.95$ are well separated from the $v_p>1$ case. This indicates the non-Gaussianity of the data, as we will further discuss in Sec.~\ref{apx_backtest}. 
 
We then consider the sample variance effect discussed in Sec.~\ref{1.D} and understand how the underlying stochasticity of the GWB changes the prediction.  

\subsection{Forecast with sample variance}
To take into account the sample variance derived in Sec.~\ref{1.D} in our forecast, the Fisher information in Eq.~\eqref{fisher_def} needs to be modified. The covariance matrix of the measurements, Eq.~\eqref{eq,covariancematrix}, now contains an additional term, $\sigma_{SV}^2$, that depends on $v_p$, making it a diagonal of $\sigma_{ij}^2 + \sigma_{SV}^2(\xi_{ij}, v_p) \equiv (\sigma_{\rm{tot}}^2)_{ij}$ terms. Using the definition in Eq.~\eqref{def_fisher_theory}, we obtain the modified Fisher information, 
\begin{equation}
    \label{def_fisher_w_SV}
    \mathcal{F}_{SV}=\sum_{\langle i,j\rangle}\left[\frac{1}{(\sigma_{\rm {tot}}^2)_{ij}}\left(\frac{\partial\Gamma(\xi_{ij},v_p)}{\partial v_p}\bigg|_{v_{p}=v_{p}^{fid}}\right)^2+\frac{1}{2(\sigma_{\rm {tot}}^2)_{ij}^2}\left(\frac{\partial \sigma_{SV}^2(\xi_{ij},v_p)}{\partial v_p}\bigg|_{v_{p}=v_{p}^{fid}}\right)^2\right]\ .
\end{equation}
\begin{figure}[h!]
    \centering
    \includegraphics[width=1\linewidth]{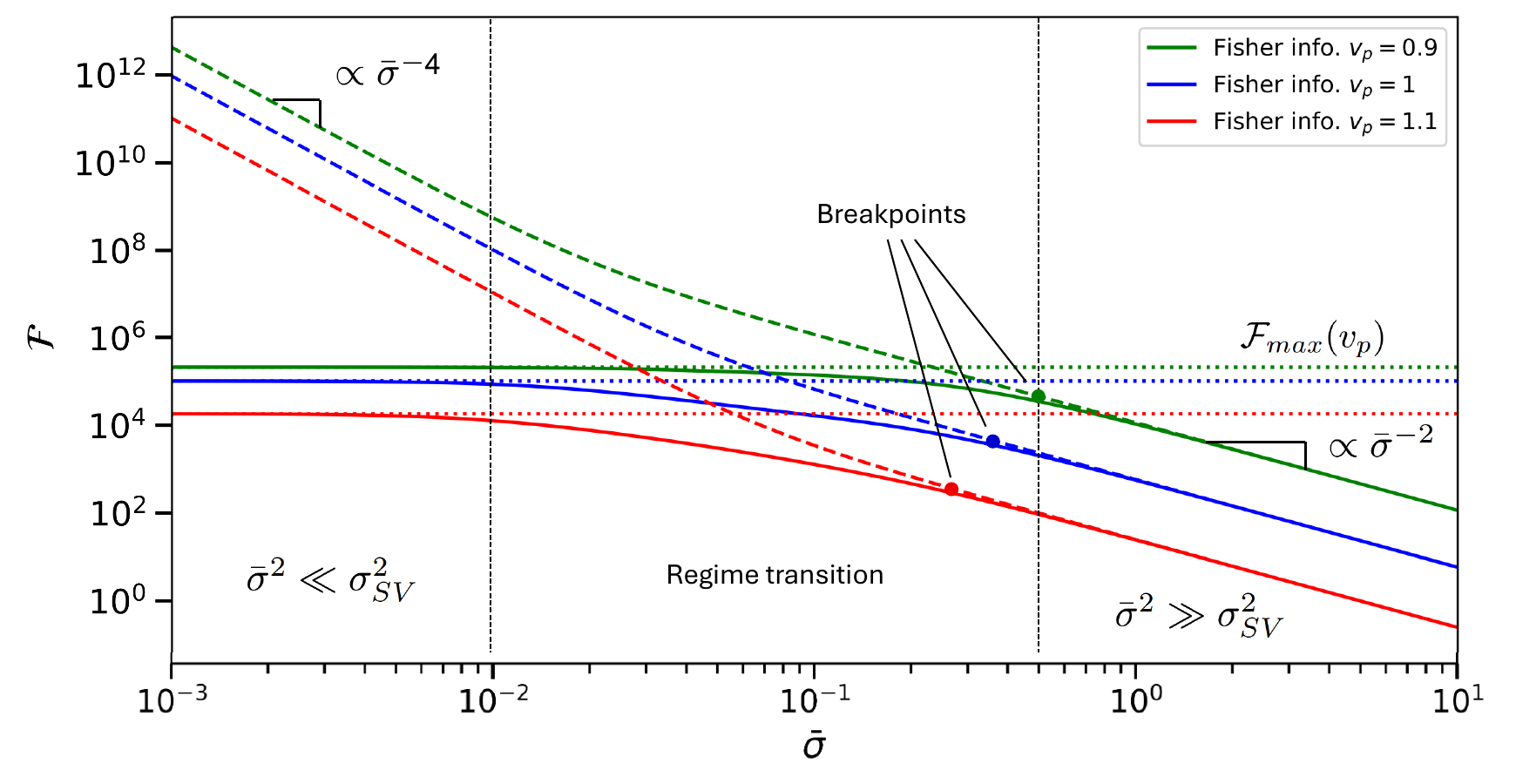}
    \caption{Total Fisher information in Eq.~\eqref{def_fisher_w_SV} as a function of the average uncertainty $\bar\sigma$ for $67$ uniformly distributed pulsars and three phase velocities, $v_p = 1.1, 1, 0.9$. Solid lines show the full Fisher information; dashed lines show the result in Eq.~\eqref{def_fisher_w_SV} obtained by approximating $\sigma_{\rm tot} \approx \bar{\sigma}$; and dotted lines show the maximum value in the limit $\sigma_{\rm SV} \gg \bar{\sigma}$. The black dotted lines separate the different approximation regimes, 
    and the breakpoints indicate the threshold where sample variance cannot be neglected. We see that the $\bar{\sigma}^{-4}$ scaling is never reached. 
    }
    \label{Fig:Fisher_info_w_SV}
\end{figure}

From Fig.~\eqref{Fig_sample_variance}, we see that the order of magnitude of the sample variance is $\sigma_{SV} \sim \mathcal{O}(0.1) \ll \bar\sigma = 1.3$, which makes $\sigma_{SV}$ small compared to the current precision. 
In this regime, we can approximate the total covariance as $(\sigma_{\rm{tot}}^2)_{ij} \approx \sigma_{ij}^2 = \bar\sigma^2$, and neglect the sample variance effect. Since the second term in Eq.~\eqref{def_fisher_w_SV} is higher order in $1/(\sigma_{\rm{tot}}^2)_{ij}$, we can drop it, and the Fisher information then reduces to Eq.~\eqref{fisher_def}, $\mathcal{F}\propto \bar \sigma^{-2}$. By contrast, in the opposite limit where the sample variance dominates the noise, $\sigma_{SV} \gg \sigma_{ij}$, we have $(\sigma_{\rm{tot}}^2)_{ij} \approx \sigma_{SV}^2$, and the Fisher information reaches a plateau as shown in Fig.~\eqref{Fig:Fisher_info_w_SV}. This limits the achievable sensitivity with PTA measurements, assuming the sample variance is constant in time \footnote{Recent work~\cite{Allen:2024uqs} develops a new strategy to reconstruct the Hellings-Downs curve, which reduces the sample variance as the observation time increases; however, applying this approach to modified gravity is beyond the scope of this paper.}.

\begin{figure}
\centering
\includegraphics[width=.9\linewidth]{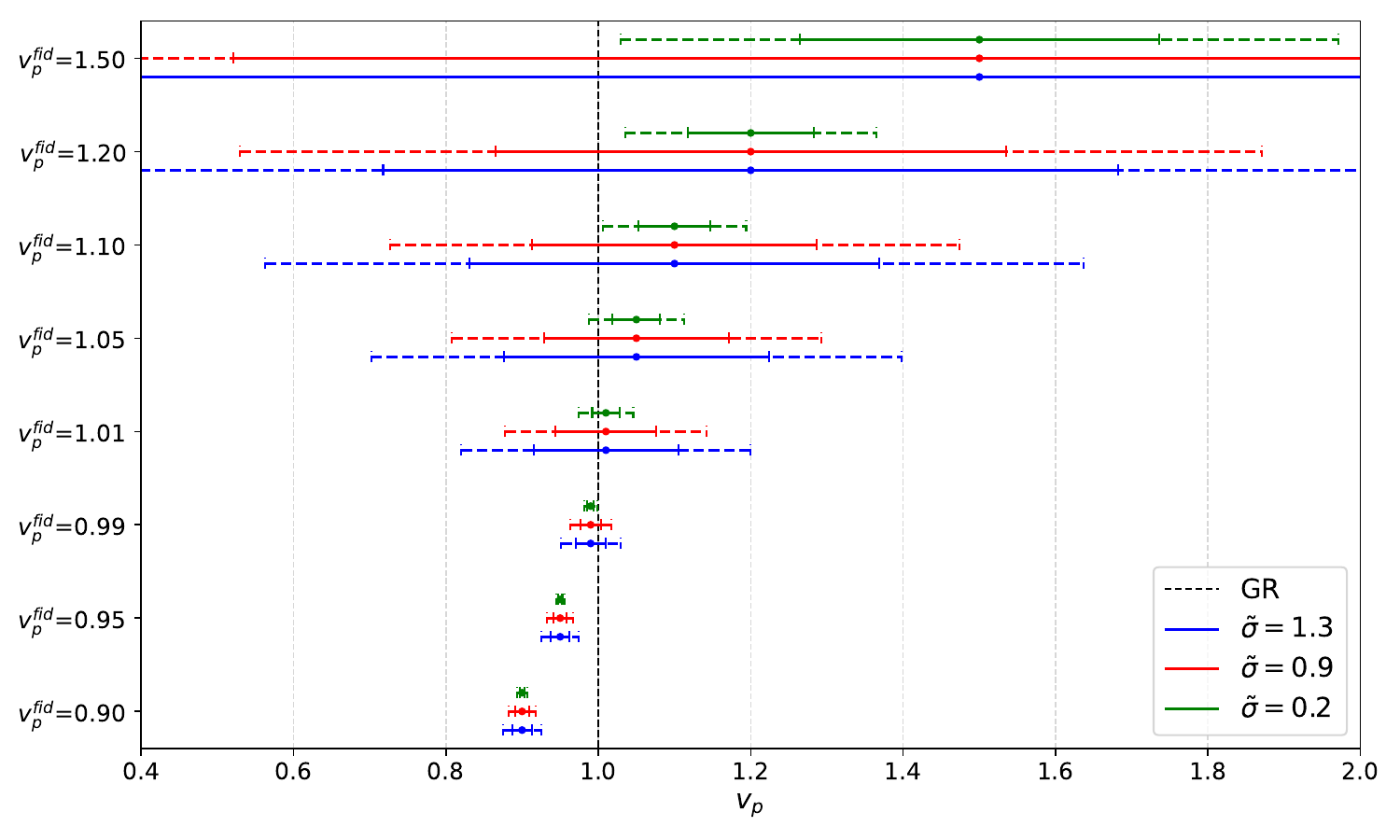}
    \caption{Whisker plot of the forecast confidence levels for different $v_p^{fid}$ and uncertainties $\tilde\sigma$, including the sample variance effect described in Sec.~\ref{1.D}. The $1\sigma$ (solid lines) and $2\sigma$ (dashed lines) are shown for comparison to the GR case ($v_p=1$, dashed black line). For large $v_p^{fid}$ and $\tilde\sigma$, the error bars exceed the window width. The results are slightly modified compared to  Fig.~\eqref{whisk_fcst} for $v_p\geq1.1$ and not even visible for $v_p<1.1$.}
    \label{fig:whiskerwithSV}
\end{figure}

Taking into account the sample variance, the required sensitivity in Table.~\eqref{tab_sigma_req} becomes Table.~\eqref{tab_sigma_req_w_SV}. Values in parentheses are those from Table.~\eqref{tab_sigma_req} for comparison. As expected, when $\tilde\sigma$ is large, the sample variance has little impact on the forecast. However, once $\sigma_{SV}$ is included, it imposes a fundamental theoretical limit on the distinguishability between GR and modified gravity with phase velocities in the range $0.99 <v_p< 1.05$, at the $3\sigma$ level. The whisker plot as a comparison to Fig.~\eqref{whisk_fcst} is shown in Fig.~\eqref{fig:whiskerwithSV}.

\begin{table}[h!]
\centering
\renewcommand{\arraystretch}{1}
\begin{tabular}{|c||c|c|c||c|c|c|c|c|}
\hline
 Deviations from GR (\%)& $-10\%$ & $-5\%$ & $-1\%$ & $+1\%$ & $+5\%$ & $+10\%$ & $+20\%$ & $+50\%$ \\
\hline
\multirow{2}{*}{$1\sigma$} 
 & \cellcolor{green!25}(10.524) & \cellcolor{green!25}(5.333) & (0.677) & (0.138) & (0.376) & (0.486) & (0.541) & (0.462) \\
 & \cellcolor{green!25}10.521 & \cellcolor{green!25}5.323 & 0.644 & 0.077 & 0.352 & 0.471 & 0.531 & 0.453 \\
\hline
\multirow{2}{*}{$2\sigma$} 
 & \cellcolor{green!25}(5.262) & \cellcolor{green!25}(2.666) & (0.339) & (0.069) & (0.188) & (0.243) & (0.270) & (0.231) \\
 & \cellcolor{green!25}5.255 & \cellcolor{green!25}2.654 & 0.270 & 0.018 & 0.148 & 0.216 & 0.251 & 0.214 \\
\hline
\multirow{2}{*}{$3\sigma$} 
 & \cellcolor{green!25}(3.508) & \cellcolor{green!25}(1.778) & (0.226) & (0.046) & (0.125) & (0.162) & (0.180) & (0.154) \\
 & \cellcolor{green!25}3.497 & \cellcolor{green!25}1.761 & 0.107 & ---   & 0.080 & 0.127 & 0.153 & 0.131 \\
\hline
\multirow{2}{*}{$5\sigma$} 
 & \cellcolor{green!25}(2.105) & (1.067) & (0.135) & (0.028) & (0.075) & (0.097) & (0.108) & (0.092)\\
 & \cellcolor{green!25}2.087 & 1.041 & ---   & ---   & 0.032 & 0.057 & 0.074 & 0.062 \\
\hline
\end{tabular}
\caption{Maximum $\tilde\sigma$ values required to distinguish different phase velocity deviations from GR at $1,2,3$ and $5\sigma$ confidence levels, taking sample variance into account and using the Fisher information defined in Eq.~\eqref{def_fisher_w_SV}. The required $\tilde\sigma$ from Table.~\eqref{tab_sigma_req} (without sample variance) are shown in parentheses for comparison. The shown percentages correspond to a phase velocity deviation. A $+10\%$ deviation corresponds to $v_p^{fid}=1.1$ and a $-10\%$ to $v_p^{fid}=0.9$. The three bars at the bottom of the table indicate that the corresponding deviations are beyond reach with this value of sample variance. Values greater than $\tilde\sigma=1.3$ computed from NANOGrav 15yr dataset~\cite{Agazie_2023} are highlighted in green.}
\label{tab_sigma_req_w_SV}
\end{table}
 
We have now calculated the required values of $\tilde\sigma$ needed to detect a given deviation from GR at a specified confidence level, both with and without including sample variance. To estimate how many years of observation are required for a potential detection, we need to relate $\tilde\sigma$ to physical parameters, such as the total observation time, the time resolution of the telescopes, and the number of observed pulsars, and model how these quantities are expected to evolve in the coming years. 
 
\section{Mock-data analysis and validation of the Fisher forecast}
\label{apx_backtest}

The Fisher forecast described in Sec.~\ref{sec2.A} assumes that the likelihood of the measured phase velocity is approximately Gaussian around the fiducial value $v_p^{\mathrm{fid}}$, with standard deviation $\sigma_v$. This assumption may break down because the log-likelihood defined in Eq.~\eqref{loglike} can be strongly sensitive to variations of the phase velocity. In particular, large noise amplitudes in the data points (corresponding to large uncertainties $\sigma_{ij}$) may lead to non-Gaussian likelihoods for the inferred phase velocity.
 
To assess the reliability of the Fisher forecast and to obtain empirical constraints from mock realizations, we perform a mock likelihood analysis using simulated data. For a given fiducial velocity $v_p^{ fid}$, we generate synthetic correlation measurements
\begin{equation}
    \rho_{ij}=\Gamma\left(\xi_{ij}, v_p^{fid}\right) + \epsilon_{ij}\ ,
\end{equation}
where $\epsilon_{ij}\sim \mathcal{N}\left(0 ,\sigma_{ij}^2\right)$ is the Gaussian distribution centered at $0$ with width $\sigma_{ij}^2$, 
and the distribution of $\xi_{ij}$ follows the probability density of Eq.~\eqref{pdf_xi}. For simplicity, we assume a reduced constant $\sigma_{ij}=\tilde\sigma$ defined in Eq.~\eqref{red_sigma}. For each realization, we compute $\chi^2(v_p)\equiv-2\log\mathcal{L}(v_p)$ using Eq.~\eqref{loglike} and infer the value of $v_p$ that minimizes $\chi^2$. 
 Repeating this procedure for a large number of realizations ($N\geq 5000$), we obtain the distribution of recovered phase velocities. From this distribution we estimate the uncertainty $\sigma_v^{\mathrm{back}}$
using the $16^{\rm th}$ and $84^{\rm th}$ percentiles, corresponding to the central $68\%$ interval.

\begin{figure}[h]
    \centering  \includegraphics[width=1\linewidth]{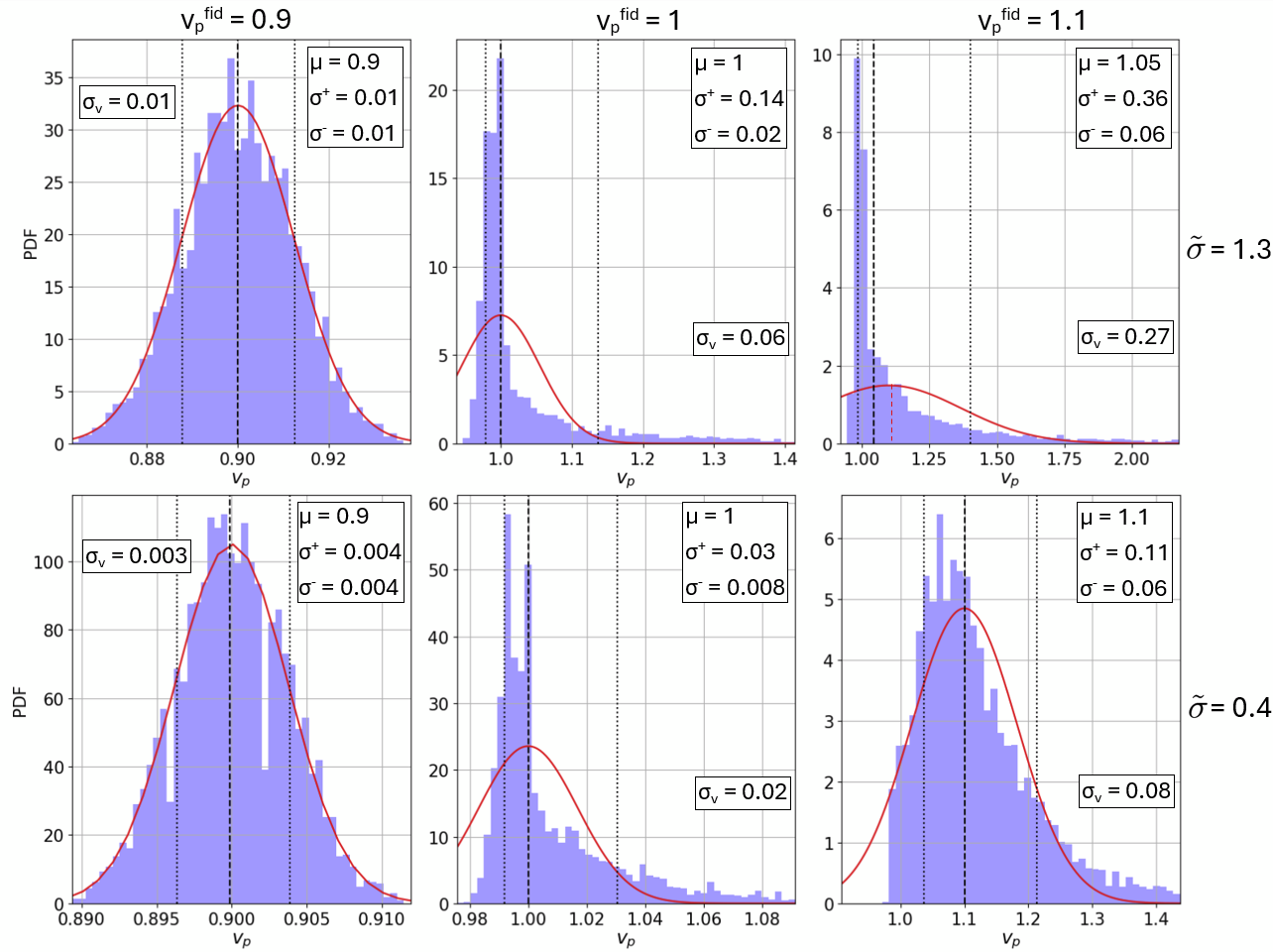}
    \caption{Posteriors of the optimal phase velocities for $v_p^{fid}\in\{0.9,1,1.1\}$ (from left to right) and standard deviation $\tilde\sigma\in\{1.3,0.4\}$ (from top to bottom). For each distribution, $\mu$ represents the $50^\text{th}$ percentile and $\sigma^+$, $\sigma^-$ the $16^\text{th}$ and $84^\text{th}$ percentiles distance from $\mu$ respectively, so that $68\%$ of the data lies in the range $[\mu-\sigma^-,\mu+\sigma^+]$. The associated forecast $\sigma_v$ and the corresponding Gaussian $\mathcal{N}\left(v_p^{fid},\sigma_v^2\right)$ (red curve) are also displayed for comparison. Note that the $x$- and $y$-axis scales are different in each plot.}
    \label{post_backtest}
\end{figure}

This mock-data analysis probes two related aspects of the Fisher approximation. First, it tests whether the estimator of the phase velocity correctly recovers the fiducial value $v_p^{fid}$. Second, it allows us to examine whether non-Gaussian features of the likelihood affect the comparison with the GR prediction $v_p=1$, which serves as a null test. While deviations from Gaussianity may degrade the accuracy of parameter recovery, they can also affect the comparison with the GR prediction, since the significance of agreement or disagreement with $v_p=1$ depends on the actual likelihood shape.

\begin{figure}[h]
    \centering   \includegraphics[width=.9\linewidth]{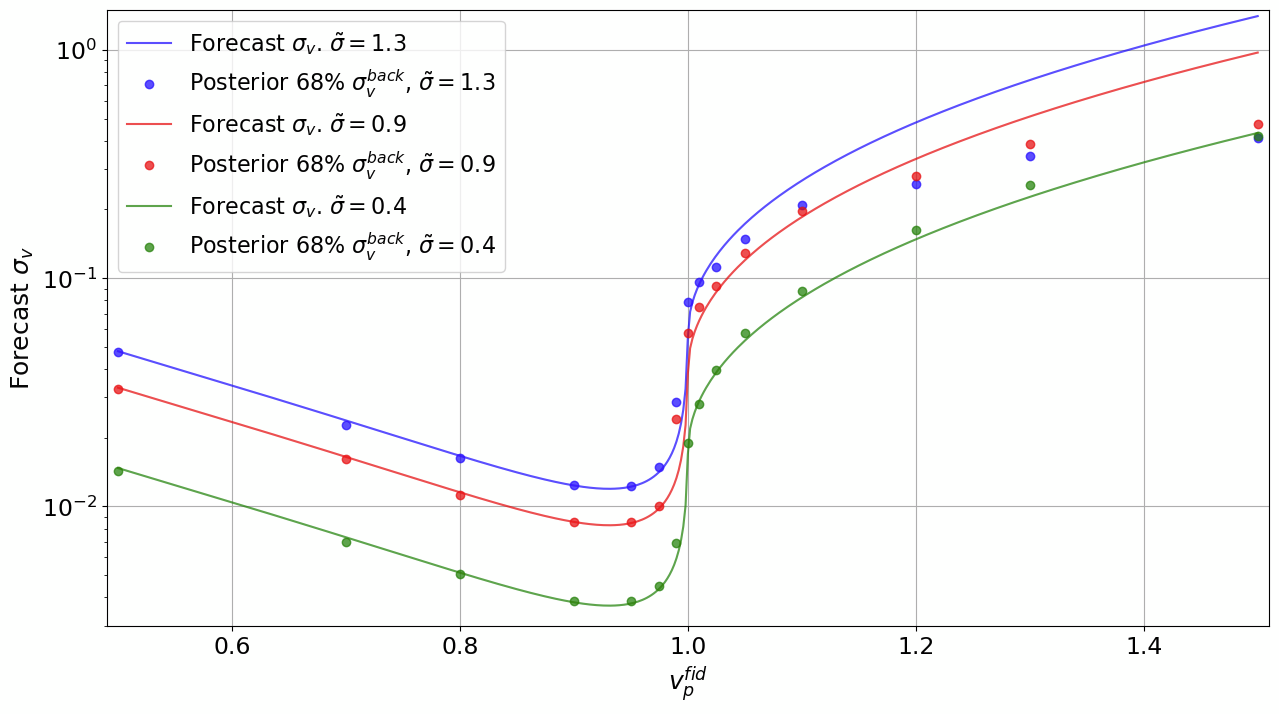}
    \caption{Comparison between the forecast $\sigma_v$ (log-scaled) from the Fisher formalism and the mock-data uncertainty $\sigma_v^{back}$, as a function of $v_p^{fid}$. The blue, red, and green curves and dots are associated with the correlation uncertainties $\tilde\sigma=\{1.3,0.9,0.4\}$, respectively. The dots represent the mean of $\sigma^-$ and $\sigma^+$, which correspond to the $16^\text{th}$ and $84^\text{th}$ percentiles distance from the median $\mu$ of the distribution, so that $68\%$ of the data lies in the range $[\mu-\sigma^-,\mu+\sigma^+]$. The blue and green dots overlap at $v_p^{fid}=1.5$.}
    \label{comp_backtest_forecast}
\end{figure}

The results of the mock-data analysis are shown in Fig.~\eqref{post_backtest} and Fig.~\eqref{comp_backtest_forecast}, which compare the Fisher forecast with the results obtained from the mock realizations. Each optimal phase velocity in the distribution can be interpreted as a realization of the measurement of the phase velocity from mock data $\rho_{ij}$, corresponding to an underlying phase velocity $v_p^{fid}$. 

Regarding the recovery of the phase velocity, the Fisher forecast provides a good approximation in most cases. This can be seen in Fig.~\eqref{post_backtest}, where we show the posteriors of the optimal phase velocities for some values of $v_p^{fid}$, with the histograms showing the mock-data analysis (the dotted line shows $\mu$, the $50^\text{th}$ percentile, and $\sigma^{-}$, $\sigma^{+}$ the $16^\text{th}$ and $84^\text{th}$ percentiles respectively, represented by the dotted lines), and the red curve a Gaussian distribution centered at $v_p^{fid}$ with standard deviation $\sigma_v$. For $v_p^{fid}<1$, the distributions of recovered phase velocities are close to Gaussian, and the Fisher forecast predicts both the velocity and its uncertainty well. For $v_p^{fid}\geq 1$, the posteriors exhibit marked right-tailed behavior. Although in most cases $v_p$ is still recovered, the inferred uncertainties can differ significantly between the Gaussian and non-Gaussian cases. As seen in the top right panel, this is not always the case, and even the recovered velocities can differ between the Gaussian and non-Gaussian assumptions. This effect diminishes as $\tilde\sigma$ decreases, since the posteriors for $v_p^{fid}>1$ approach Gaussian distributions.
 
\begin{figure}
    \centering
\includegraphics[width=.9\linewidth]{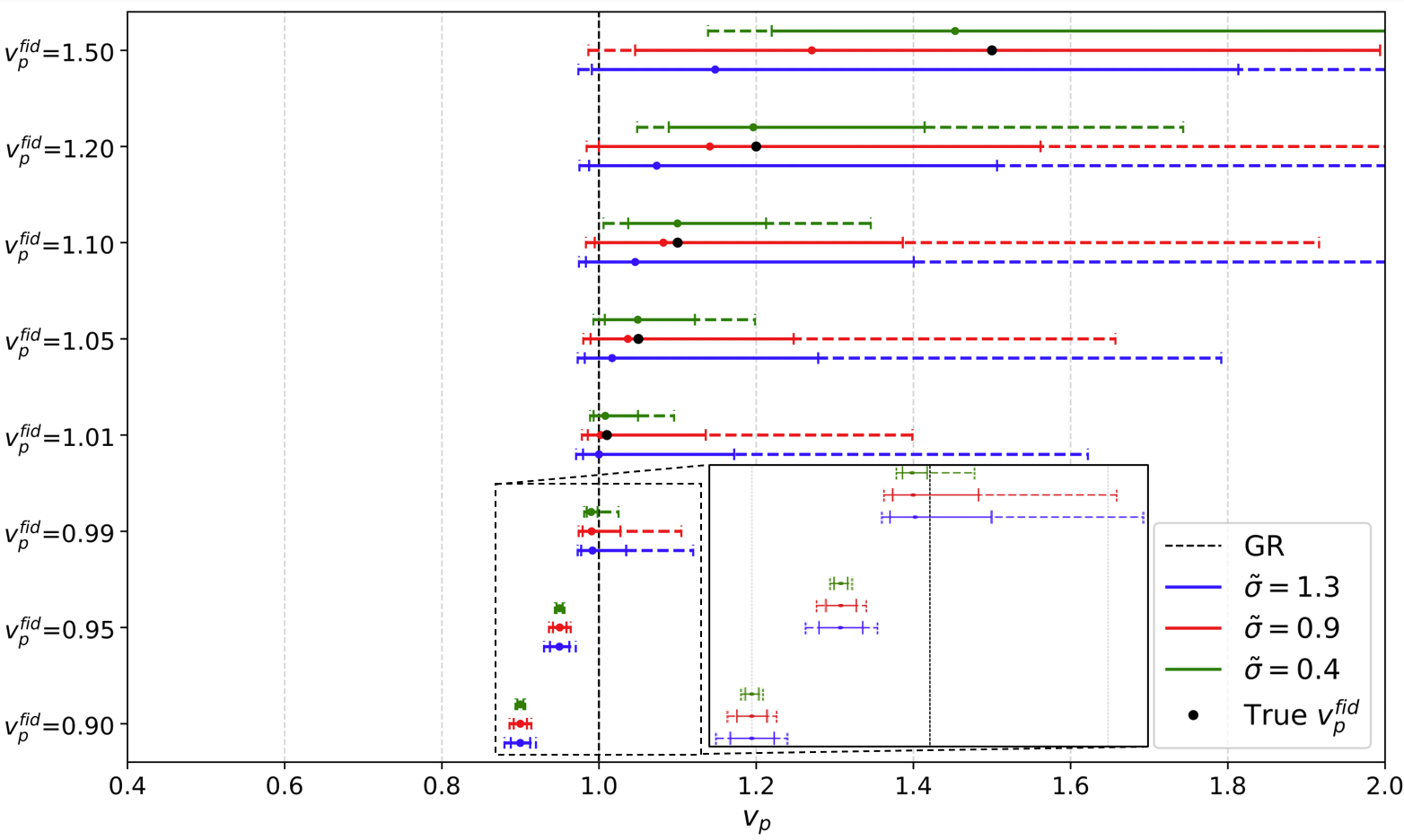}
    \caption{Whisker plot of the mock-data distribution confidence levels for different $v_p^{fid}$ and uncertainties $\tilde\sigma$. The $1\sigma$ (solid lines) and $2\sigma$ (dashed lines) are shown for comparison to the GR case ($v_p=1$, dashed black line). Note that the green lines correspond to $\tilde\sigma=0.4$, in contrast to Fig.~\eqref{whisk_fcst} and \eqref{fig:whiskerwithSV} (where $\tilde\sigma=0.2$). The backtest distributions exhibit large tails and offset means for $v_p>1$. The offset and the tails tend to vanish when $\tilde\sigma$ decreases. Fiducial phase velocities above $1.1$ start to be distinguishable at $2\sigma$ level for $\tilde\sigma=0.4$.}
    \label{whisk_backtest}
\end{figure}

These differences can be quantified in Fig.~\eqref{comp_backtest_forecast}, which compares the Fisher uncertainty $\sigma_v$ (solid lines) with the uncertainty obtained from the mock realizations $\sigma_v^{\mathrm{mock}}$ (dots) as a function of the fiducial phase velocity. To have a single value of $\sigma_v^{back}$, we take the mean of the $16^\text{th}$ and $84^\text{th}$ percentiles ($\sigma^-$ and $\sigma^+$, respectively) from the median $\mu$ of the distribution. As $\tilde\sigma$ decreases, the mock-data uncertainty approaches $\sigma_v$ because the posteriors approach Gaussian distributions. For $v_p^{fid}<1$, the dots and solid lines are aligned, which is consistent with the results of Fig.~\eqref{post_backtest} where posteriors are Gaussian. For $v_p^{fid}>1$ and large $\tilde\sigma$, the forecast systematically overestimates $\sigma_v$ for large fiducial phase velocities, and the error shrinks as $\tilde\sigma$ decreases.

This is illustrated in Fig.~\eqref{whisk_backtest}, where we assess the impact of the Gaussian approximation used in the Fisher formalism (i.e., assuming a parabolic $\chi^2$ near its minimum) by presenting a whisker representation of the mock-data distributions. This figure should be compared to Fig.~\eqref{whisk_fcst}. Deviations from Gaussianity can, for instance, slightly shift the distribution, as illustrated, for example, by black dots on red lines for $v_p>1$, as difference between the median of the confidence interval and the true value $v_p^{ fid}$. In some cases, this shift moves the distribution closer to the GR value, effectively degrading the forecasted separation from GR.

The comparison with the GR prediction $v_p=1$, however, is more sensitive to the Gaussian assumption. As shown in Fig.~\eqref{whisk_backtest}, for $v_p^{fid}>1$ the Gaussian uncertainty overestimates the error on the left side and underestimates it on the right side due to the tail in the mock-data distributions. The value $v_p = 1$, corresponding to GR, serves as a threshold for the distributions: to the left of this value, the distributions are approximately Gaussian, whereas to the right they become noticeably non-Gaussian for current precision. 
Although the long tail is an important feature of the mock-data distributions, the main consequence for the forecasts at $v_p > 1$ is that the Fisher approximation overestimates the uncertainty on the left side compared to the mock-data results. As a result, cases that appear compatible with GR in the Fisher forecast may in fact be distinguishable in the mock realizations, where the left-side uncertainty is smaller. In this sense, the Fisher forecast can lead to overly conservative null tests for $v_p > 1$, and a mock analysis, such as the one performed here, or a full MCMC analysis, becomes necessary.
An exception occurs for $v_p^{fid} = 0.99$, where the tail enlarges the right-side uncertainty and increases the apparent compatibility with GR, making the Fisher forecast optimistic in this case.

We conclude from this mock-data analysis that, although the Fisher forecast provides a useful Gaussian approximation for the recovery of the phase velocity, it also reveals systematic differences once the likelihood becomes non-Gaussian. In particular, non-Gaussian features of the likelihood, which arise from the sensitivity of the ORF to the phase velocity, can affect the interpretation of the results when comparing with the GR prediction $v_p=1$. The uncertainty estimated from the mock realizations approaches the Fisher prediction $\sigma_v$ as $\tilde\sigma$ decreases and the distributions become closer to Gaussian.
However, for $v_p^{fid} > 1$, the distributions develop pronounced right tails and can also exhibit small shifts of their central value, introducing an asymmetry in the uncertainties. In particular, the Fisher approximation tends to overestimate the uncertainty on the left side of the distribution, which is the side relevant for assessing the overlap with the GR value. As a consequence, the Fisher forecast can lead to overly conservative null tests in this regime.
While the Fisher approach remains useful for estimating the order of magnitude of the precision required to detect deviations from GR, a more accurate treatment of the likelihood, such as the Monte Carlo mock analysis performed here or a full MCMC exploration, is required for a reliable assessment of the compatibility with GR.

This mock-data analysis should be regarded as an intermediate step aimed at probing the validity of the Gaussian approximation underlying the Fisher formalism. In particular, it allows us to compare the Fisher prediction for the uncertainty on $v_p$ with the sampling distribution obtained from repeated mock realizations. While this procedure does not replace a full likelihood exploration, it provides a useful diagnostic of potential non-Gaussian features in the likelihood and their impact on the inferred uncertainties.

\section{Prospects for deviation detection}
\label{Sec3}
In this section, we estimate the number of observation years required to find deviations from GR, based on the sensitivity of current PTA collaborations. Following the results described in~\cite{Anholm_2009, babak2024forecastingsensitivitypulsartiming}, we relate the reduced uncertainty $\tilde{\sigma}$ to the observation time. We then explore several scenarios, corresponding to different assumptions about observational strategies and millisecond pulsar discoveries, to assess how future datasets might enable us to distinguish specific deviations from GR.
 
We begin with the existing telescopes and PTA systems to assess whether observations over the next decade may reveal any deviations. To do this, we focus on the NANOGrav collaboration, which has observed 67 pulsars, and we take the current precision to be $\tilde{\sigma}_1 = 1.3$ and $\tilde{\sigma}_2 = 0.9$, where the latter better captures the uncertainty in the $v_p > 1$ regime, as discussed in Fig.~\eqref{Comp_NG_rand}. To achieve the desired precision in Table.~\eqref{tab_sigma_req} and \eqref{tab_sigma_req_w_SV}, we expect that longer observation times will improve the sensitivity for each individual pulsar pair, $\sigma_{ij}$, and that additional millisecond pulsar discoveries will further enlarge the array.

\begin{figure}[h!]
    \centering
  \includegraphics[width=\linewidth]{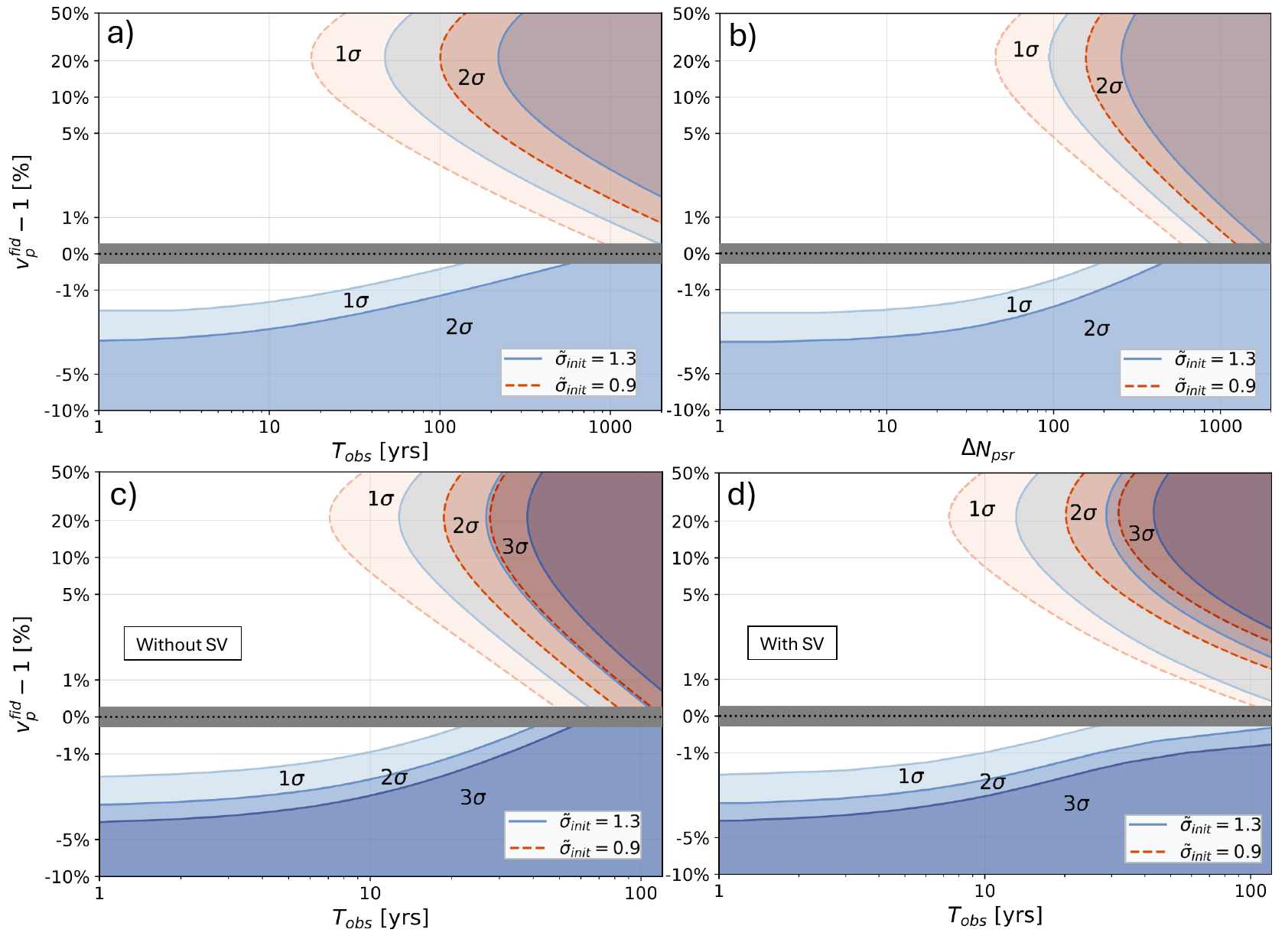}
    \caption{Log–log plots of the potential deviation from GR in the gravitational-wave phase velocity (in \%) as a function of the required observation time (in years) or number of pulsars needed for detectability with NANOGrav’s PTA. The blue shaded regions correspond to the conservative initial $\tilde\sigma_1=1.3$ and the red dashed shaded regions to $\tilde\sigma_2=0.9$, when focusing on the case $v_p>1$ as in Fig.~\eqref{Comp_NG_rand}. From lighter to darker regions, the $1\sigma$, $2\sigma$ (and $3\sigma$) confidence levels are shown. \textbf{a)} Detectability in Scenario 1. We only increase observation time, keeping the same number of pulsars in the dataset. \textbf{b)} Detectability in Scenario 2. We only increase the number of pulsars and forget about the time dependence of $\tilde\sigma$. \textbf{c)} Detectability in Scenario 3. Two effects are combined to improve the initial precisions $\tilde{\sigma}_1$ and $\tilde{\sigma}_2$. Increasing the observation time reduces the uncertainty, and we extend the dataset with 6 additional pulsars per year, matching the rate of new pulsar discoveries in the NANOGrav dataset. We use the total $\tilde{\sigma}$ time scaling from Eq.~\eqref{time_scale_sigma}. \textbf{d)} Same results including the sample variance effect described in Sec.~\ref{1.D}.
}
    \label{T_obs_needed_all}
\end{figure}

We recall from the usual optimal statistic derivation~\cite{Anholm_2009, babak2024forecastingsensitivitypulsartiming, Siemens_2013} that the uncertainty for a pulsar pair, $\sigma_{ij}$, is inversely proportional to the square root of the common observation time of the two pulsars, $T_{ij}=\min(T_i,T_j)$. After an additional observation period of duration $\Delta T$, $\sigma_{ij}$ reduces to
\begin{equation}
\label{time_scale_sig_ij}
    \sigma_{ij}(T_{ij}+\Delta T)=\frac{\sigma_{ij}(T_{ij})}{\sqrt{1+\frac{\Delta T}{T_{ij}}}}\ .
\end{equation}
Alongside the definition of $\bar\sigma$ in Eq.~\eqref{eq,defbarsigma}, we want to define a quantity, $T_0$, that represents the common observation time for the pulsar sets, and study the overall time scaling of $\tilde \sigma$ to make forecasts. For the following definition,
\begin{eqnarray}
\label{eq,defT0}
   T_0 \equiv  \sum_{<ij>}\frac{1}{\sigma_{ij}^2(T_{ij})}\bigg/{\sum_{<ij> } \frac{1}{T_{ij}\,\sigma_{ij}^2(T_{ij})}} \approx 10~\text{yrs} \ ,  
\end{eqnarray}
where we use NANOGrav’s 15-year result to estimate $T_0$~\cite{Agazie_2023, the_nanograv_collaboration_2025_16051178}, one can find the overall time scaling behavior of uncertainty defined in Eq.~\eqref{red_sigma} as 
\begin{equation}
\label{sigma_vs_time}
    \tilde\sigma(T_0+\Delta T) = \frac{67}{\sqrt{\sum  \frac{1}{\sigma_{ij} ^2(T_{ij}+\Delta T ) } }} = \frac{67}{\sqrt{\sum  \frac{1}{\sigma_{ij} ^2(T_{ij} ) }  \left(1+ \frac{\Delta T \sum 1/(\sigma_{ij}^2 T_{ij}) }{\sum 1/\sigma_{ij}^2 }\right)} } =\frac{\tilde \sigma(T_0)}{\sqrt{1+\frac{\Delta T}{T_0}}}\ ,
\end{equation}
without taking into account the newly observed pulsars. For the same pulsar sets, the common precision $\bar\sigma$ also precludes the same time scaling. Fig.~\eqref{T_obs_needed_all}.a shows the scaling for two different initial values, $\tilde\sigma = 1.3$ and $0.9$. One can see that $\sim25$ years are required to reach a $10\%$ deviation at the $1\sigma$ level when assuming the initial uncertainty to be $\tilde \sigma = 0.9$ as discussed in Fig.~\eqref{whisk_fcst}.a.

As a comparison, in Fig.~\eqref{T_obs_needed_all}.b we study the effect of the pulsar number on $\tilde\sigma$ by fixing the precision $\bar\sigma$. Since we already have $67$ pulsars in NANOGrav, adding $100$ pulsars improves the reduced uncertainty by only a factor of $167/67\approx 2.5$. 

To study the improvement in precision from increasing the number of observed pulsars over time, we use the pulsar discovery rate reported in~\cite{Agazie_2023}, which suggests that roughly six new pulsars are added to the data set each year. Using this rate, $r$, and starting from the initial number $N_0 = 67$, the reduced uncertainty defined in Eq.~\eqref{red_sigma} can be expressed as, 
\begin{equation}
    \tilde\sigma(T_0+\Delta T)=\sqrt{ P(N_0 )} \Bigg/ \sqrt{\sum_{<ij>\in a}\frac{1}{\sigma_{ij}^2(T_{ij}+\Delta T)}+\sum_{m=0}^{\Delta T-1}\sum_{ <ij>\in b_m  }\frac{1}{\sigma_{ij}^2(T_1+m)}}\ ,
\end{equation}  
 where $P(N)=N(N-1)/2$ denotes the number of pulsar pairs for $N$ pulsars. The precision in the denominator contains two parts: subset $a$ represents those pulsars that are already in the current data set, while $b_m$ denotes the pairs formed by the newly added pulsars between year $m$ and $m+1$, including the pairs among themselves and with the existing pulsars. Here, $T_1\approx 3$ years is the common observation time for a new pulsar to be added to the data set according to~\cite{Agazie_2023}. Assuming the newly added pulsar pairs will reach the same precision $\bar\sigma(T_0)$ at time $T_0$, we can represent the second term in the denominator as 
\begin{equation}
     \sum_{<ij>\in b_m}\frac{1}{\sigma_{ij}^2(T_1+m)}\approx \frac{P(N_0+rm+r)-P(N_0+rm) }{\bar\sigma^2(T_1+m)} \ ,
\end{equation}
where the numerator captures the number of pulsar pairs in the set $b_m$. 
Using the time scaling of $\bar\sigma (T_1+m)\sim \bar \sigma (T_0)\sqrt{T_0}/\sqrt{T_1+m}$ for these pulsar sets, we end up with
\begin{eqnarray}
\label{time_scale_sigma}
\tilde\sigma(T_0+\Delta T)&=& \tilde \sigma(T_0)\sqrt{\frac{P(N_0)}{\left(1+\frac{\Delta T}{T_0}\right)P(N_0)+\sum_{k=0}^{\Delta T-1}\frac{T_1+k}{T_0}\left[P(N_0+rk+r)-P(N_0+rk)\right]}}\nonumber\\
    &\approx &\frac{\tilde\sigma(T_0)}{\sqrt{1+\frac{\Delta T}{T_0}\left[1+\frac{r\big(T_1-\frac{1}{2}\big)}{N_0}\left(2+\frac{r\Delta T}{N_0}\right)+\frac{r\Delta T}{N_0}+\frac{2}{3}\left(\frac{r\Delta T}{N_0}\right)^2\right]}}\ ,    
\end{eqnarray}
where we take the limit $N_0 \gg r > 1$. The preliminary observation time $T_1$ only plays a role when $\Delta T \lesssim T_1$, with larger values of $T_1$ leading to a faster decrease of $\tilde\sigma$. When $\Delta T > T_1$, the term proportional to $\Delta T^2$ dominates the sum in the denominator.

We show this forecast in Fig.~\eqref{T_obs_needed_all}.c without including the sample variance effect, and in panel d with the sample variance effect included. The optimistic forecast is indeed improved compared to the Fig.~\eqref{T_obs_needed_all}.a. It can be seen that for $v_p<1$, the sensitivity to distinguish a -1\% deviation from GR could be reached at $3\sigma$ level within 30 years of observation under the optimistic forecast. For $v_p>1$, in the most favorable scenario (orange contours), a $+10\%$ deviation could be detected at the $3\sigma$ level within about 30 years, whereas such sensitivity is not reached within 1000 years in Fig.~\eqref{T_obs_needed_all}.a. In the same scenario, detecting a $+1\%$ deviation at the $2\sigma$ level would require roughly 60 years.

An interesting feature in Fig.~\eqref{T_obs_needed_all} is the observation that a +20\% deviation is more easily distinguished than a +50\% deviation. This counter-intuitive effect arises from the phase velocity dependence of the ORF, which can be seen from the definition of $\sigma-$ distance in Eq.~\eqref{def_sigma_dist}. For large $v_p$ (typically $v_p \gtrsim 1.5$), the octupole and higher-order multipole modes are exponentially suppressed in the ORF's spherical harmonic decomposition, reducing it to the quadrupole moment. In this regime, the ORF becomes nearly insensitive to $v_p$, causing the Fisher information to approach 0 and $\sigma_v$ to diverge. Although the numerator of $d_\sigma$ increases as $v_p$ increases, the denominator $\sigma_v$ diverges faster, making it harder to distinguish from general relativity. Furthermore, as the Fisher information approaches zero, the Fisher formalism breaks down because it no longer accurately represents the curvature of the $\chi^2$ statistic (see Appendix~\ref{apx_Fisher}). Specifically, $\chi^2(v_p)$ becomes asymptotically flat at high velocities, failing to maintain a parabolic profile within the interval $[v_p^{fid}-\sigma_v, v_p^{fid}+\sigma_v]$.

We have provided an optimistic forecast of the future sensitivity to the fiducial phase velocity using the current NANOGrav 15-year dataset. We first studied the time scaling of the precision for pulsar pairs without including newly observed pulsars, then examined the effect of the number of pulsars assuming a fixed timing uncertainty, and finally combined these two effects to assess how newly added pulsars impact the overall uncertainty. Assuming that the PTA can add six new pulsars to the dataset per year, it is feasible to distinguish a phase velocity of $1.1$ and $0.99$ from the GR prediction at the $3\sigma$ level within 40 years, including the effect of sample variance.

\section{Conclusion and Discussion}

In this work, we predicted the sensitivity needed by PTA surveys to detect potential deviations from GR in the phase velocity of the gravitational wave background. We recalled the PTA response in the context of a modification of the dispersion relation to the tensor modes, following the results of~\cite{Liang_2023, liang2024testinggravityfrequencydependentoverlap, Liang_2021} and applied the Fisher formalism to the overlap reduction function to predict an uncertainty on the phase velocity, using a Gaussian approximation of the likelihood. We tested its validity and explored the effects of non-Gaussian behaviors on the forecast through a mock-data analysis. We found that in most cases, the Fisher forecast is a conservative approach to test for phase velocity deviations. We showed that the expected theoretical uncertainty is much higher for a superluminal phase velocity than for the subluminal case, in accordance with current PTA surveys on modified gravity~\cite{liang2024testinggravityfrequencydependentoverlap,Bernardo_2023,Wu_2024,Wang_2024}. We explored the effect of sample variance on the measured correlations and how it affects our forecast. More specifically, we have given concrete results on the optimal statistical uncertainty required to detect many potential deviations from GR, at different confidence levels and with or without considering sample variance. 

Using the time-scaling laws presented in~\cite{Anholm_2009, Siemens_2013}, we derived the time dependence of the reduced uncertainty, both without newly discovered pulsars, as shown in Eq.~\eqref{sigma_vs_time}, and including their contribution, as shown in Eq.~\eqref{time_scale_sigma}, assuming a rate of six pulsars added to the dataset each year based on NANOGrav PTA observations. 
We find that approximately thirty (forty) years of observations would be required to exclude general relativity at the $3\sigma$ level if the true phase velocity is 1.1 or 0.99 (See Fig.~\eqref{T_obs_needed_all} c(d)) without (with) sample variance effect. 

However, this optimistic forecast relies on several key assumptions, both theoretical and observational. First, we assumed an infinite distance between pulsars and Earth to obtain the analytical expression for the coefficients of the overlap reduction function in Eq.~\eqref{approx_cl}. This leads to a nonphysical divergence at $\xi = 0$, which prevents the use of the usual normalization convention, $\Gamma(\xi = 0) = 1/2$. We therefore adopt a different convention, $\Gamma(\xi = \pi) = 1/4$, throughout the paper. Second, we used a frequentist approach, relying on the maximum-likelihood estimate for the phase velocity and the Fisher forecast for our analysis. As shown in the mock-data study, the Fisher forecast provides a good estimation of the order of magnitude of $\sigma_v$, but a full MCMC exploration of the likelihood would provide a more realistic forecast, which is beyond the scope of this paper.

Besides the theoretical assumptions, we also make several observational assumptions. First, we assumed a common uncertainty, $\bar{\sigma}$, to perform the forecast (Eq.~\eqref{eq,defbarsigma}). We further assumed a uniform distribution of pulsars, as discussed in Eq.~\eqref{pdf_xi}, to obtain this uncertainty, $\bar{\sigma} = 1.3$, from the NANOGrav dataset. As shown in the left panel of Fig.~\eqref{Comp_NG_rand}, $\bar{\sigma} = 1.3$ does not fully represent the current constraining power for the case $v_p > 1$. We therefore adopt a smaller uncertainty, $\bar{\sigma} = 0.9$, to provide a more optimistic forecast in this regime. Second, to convert the required precision into the observation time, we assume that the existing pulsars in the array do not lose stability over long-term observations and that, in addition, six new millisecond pulsars will be added each year. We further assume that newly formed pulsar pairs will approach the same uncertainty, $\bar{\sigma}(T_0)$, if they are observed for a common duration $T_0$. Although the precision of individual pulsar pairs depends strongly on the specific pulsars in the array, we expect that a sufficiently large set of newly discovered pulsars will statistically approach the same uncertainty.

Given these assumptions and the feasible timescale indicated by the forecast, further study toward a more robust prediction is required. A simulation of the pulsar distribution, together
with a Bayesian analysis, would provide a better estimate of the achievable precision and the
angular distribution of the pulsars; this is left for future work. 
Another way to improve this forecast is to study the finite-distance effects between the pulsars and the Earth in more detail, although these distances are currently poorly measured in PTA observations. The sample variance intrinsic to the gravitational-wave background may also be better reconstructed using methods such as those discussed in~\cite{Allen:2024uqs} for the general-relativistic Hellings–Downs curve.

 Luckily, the ongoing International Pulsar Timing Array (IPTA) collaboration may provide improved sensitivity by combining datasets from existing PTA collaborations. The FAST telescope in the Chinese PTA and MeerKAT in South Africa also offer promising improvements for near-future observations and are expected to discover more high-quality millisecond pulsars, which will significantly improve the feasibility of the forecast, as discussed in Sec.~\ref{Sec3}. The timing precision of future facilities such as the Square Kilometre Array (SKA)~\cite{janssen2014gravitationalwaveastronomyska,babak2024forecastingsensitivitypulsartiming} is also expected to be 5--10 times better than that of the telescopes currently used by NANOGrav~\cite{Agazie_2023,joshi2022nanohertz}.

In conclusion, detecting deviations of the phase velocity from the general-relativistic prediction, as a probe of modified gravity in the nanohertz band, may be achievable within the next three to four decades. A more realistic assessment is left for future work.

\begin{acknowledgments}
We thank Kimberly Boddy, Yifan Chen, Tristan Smith and Joe Romano for useful discussions.  We thank  Yifan Chen, Reginald Christian Bernardo and Kimberly Boddy for helpful comments on the draft. This work is supported by ILANCE, CNRS/IN2P3, University of Tokyo, Japan. QL and EF are supported by World Premier International Research Center Initiative (WPI Initiative), MEXT, Japan.  
\end{acknowledgments} 

\appendix
\section{Choice of ORF normalization}
\label{apx_ORF}
When relying on the Fisher formalism for our forecast, an important issue to discuss is ORF normalization. We recall from Sec.~\ref{sec1} that the overlap reduction function is defined up to a normalization factor. In other works~\cite{Anholm_2009,Agazie_2023}, this normalization is made such that $\Gamma(0^\circ,v_p)=1/2$. We cannot use this method since $\Gamma(0^\circ,v_p)$ diverges for $v_p<1$ in the infinite distance limit. We therefore need to choose another reference angle $\theta_0$. We normalize the ORF so that 
\begin{equation}
    \Gamma(\theta_0,v_p)=\text{HD}(\theta_0)\ ,
\end{equation}
where HD is the Heilings-Downs curve function defined in Eq.~\eqref{def_HD}. The forecast should be almost invariant on the choice of this reference angle. We exclude the angles for which there exist a phase velocity such that $\Gamma(\theta_0,v_p)=0$. The range to avoid is $\theta_0\in \left[36^\circ,55^\circ\right]\cup\left[116^\circ,125^\circ\right]$. 
\begin{figure}[h]
    \centering
    \includegraphics[width=1\linewidth]{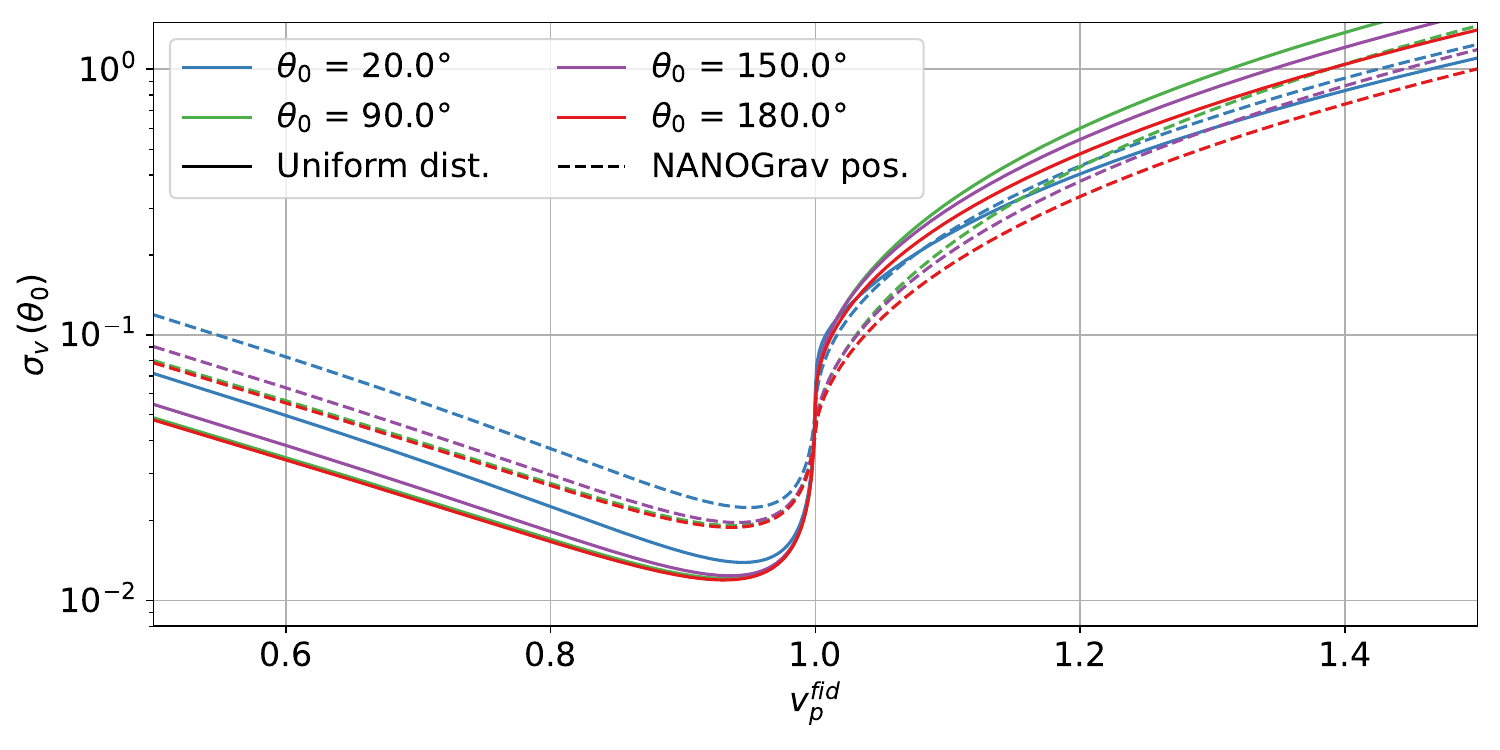}
    \caption{Forecast $\sigma_v$ (log-scaled) from the Fisher formalism as a function of $v_p^{fid}$, for different values of the reference angle $\theta_0$. The fiducial velocity ranges over $[0.5,1.5]$. The curves overlap for $v_p^{fid}\in[0.95,1.05]$. We use $\tilde\sigma=1.3$ defined in Eq.~\eqref{red_sigma} for the solid lines and NANOGrav data~\cite{the_nanograv_collaboration_2025_16051178} for the dashed lines.}
    \label{Norm_deviat}
\end{figure}
For a reference angle sufficiently far from the excluded interval, the forecast is not very sensitive to the choice of the reference angle. Fig.~\eqref{Norm_deviat} shows the forecasted $\sigma_v$ as a function of the fiducial phase velocity for different normalization choices. Given these results, we choose $\theta_0=180^\circ $ for the following reasons: the curve corresponding to $\theta_0=180^\circ$ is quite close to the three other curves and the maximal deviation in $\sigma_v$ observed with this range of fiducial velocities and reference angles is $50$\% at $v_p^{fid}=0.5$. Furthermore, at this angle, the Heilings-Downs correlation is simply $1/4$ and the Legendre polynomials $P_\ell(\cos\theta_0)$ are $(-1)^\ell$, which simplifies the computations and avoids numerical errors. Moreover, the distribution of angular separations $\xi_{ij}$ among pulsar pairs is such that significantly fewer pairs lie near $\xi_{ij} = 180^\circ$ compared to $\xi_{ij} = 90^\circ$. Since the derivative of the normalized ORF with respect to $v_p$ vanishes at the reference angle $\theta_0$, selecting $\theta_0 = 180^\circ$ results in the exclusion of fewer data points in the computation of the Fisher information than choosing any other reference angle. We therefore normalize the ORF so that $\Gamma(180^\circ,v_p)=1/4$.

\section{Fisher information and log-likelihood curvature}
\label{apx_Fisher}
The maximum likelihood estimator (MLE) is a tool used in most frequentist analyzes to determine an unknown parameter. If used to determine the true parameter $v_p^{true}$, our goal is to determine the uncertainty on this parameter $v_p^{true}$. We recall the definition of the log-likelihood from Eq.~\eqref{loglike}
\begin{eqnarray}
    &\log\mathcal{L}\,(\Gamma^{obs},v_p)=-\frac{1}{2}\left(\Gamma^{obs}-\Gamma(v_p)\right)^T\Sigma^{-1}\left(\Gamma^{obs}-\Gamma(v_p)\right)-\frac{1}{2}\log \left(2\pi\det\Sigma\right)\ ,
\end{eqnarray}
where $\Sigma$ is the covariance matrix of the measured correlations $\Gamma^{obs}$ and $\Gamma(v_p)$ is the model. When the number of pulsars $N_{psr}\gg 1$, the random variable $-2\log\mathcal{L}$ approaches the $\chi^2$ distribution. We define the new random variable
\begin{equation}
\chi^2(\Gamma^{obs},v_p)\equiv-2\log\mathcal{L}(\Gamma^{obs},v_p) \ .
\end{equation}
Maximizing the log-likelihood is equivalent to minimizing the $\chi^2$. We can Taylor-expand $\chi^2$ around the parameter $v_p^{true}$ and we restrict ourselves to the second order in $\Delta v_p\equiv v_p-v_p^{true}$
\begin{equation}
\label{taylor_exp}  \chi^2(\Gamma^{obs},v_p)=\chi^2(\Gamma^{obs},v_p^{true})+\Delta v_p\frac{\partial\chi^2}{\partial v_p}\bigg|_{v_p^{true}} +\frac{1}{2}(\Delta v_p)^2\frac{\partial^2\chi^2}{\partial v_p\,^2}\bigg|_{v_p^{true}}+\mathcal{O}\left(\Delta v_p^3\right)\ .
\end{equation}
We have
\begin{equation}
\label{sec_deriv_chi}
    \frac{\partial^2\chi^2}{\partial v_p\,^2}=-2\frac{\partial^2\log\mathcal{L}}{\partial v_p\,^2}=-2\left(\frac{1}{\mathcal{L}}\frac{\partial^2\mathcal{L}}{\partial v_p\,^2}-\left(\frac{\partial\log\mathcal{L}}{\partial v_p}\right)^2\right)\ .
\end{equation}
Since $\mathcal{L}(\Gamma^{obs},v_p)\equiv f(\Gamma^{obs}|\,v_p)$ where $f$ is the parameter-dependent probability density function of $\Gamma^{obs}$, we also have
\begin{equation}
\begin{aligned}
    \mathbf{E}\left(\frac{1}{\mathcal{L}}\frac{\partial^2\mathcal{L}}{\partial v_p\,^2}\bigg|\,v_p^{true} \right)&=\int \frac{1}{\mathcal{L}(\Gamma^{obs},v_p^{true})}\frac{\partial^2\mathcal{L}(\Gamma^{obs},v_p^{true})}{\partial v_p\,^2}f(\Gamma^{obs}|\,v_p^{true})\,d\Gamma^{obs}\\
    &=\int \frac{\partial^2\mathcal{L}}{\partial v_p\,^2}\,d\Gamma^{obs}=\frac{\partial^2}{\partial v_p\,^2}\int f(\Gamma^{obs}|\,v_p^{true})\,d\Gamma^{obs}\\
    &=\frac{\partial^2 }{\partial v_p\,^2}\,1=0\ .\\
    \end{aligned}
\end{equation}
Taking the expectation values on both sides in Eq.~\eqref{sec_deriv_chi} and using the definition of the Fisher information in Eq.~\eqref{fisher_def_theoric}, we obtain
\begin{equation}   \mathbf{E}\left(\frac{\partial^2\chi^2}{\partial v_p\,^2}\bigg|\,v_p^{true} \right)=\mathbf{E}\left(\frac{\partial^2\chi^2}{\partial v_p\,^2}\bigg|_{v_p^{true}}\right)=2\mathcal{F}(v_p^{true}) \ .
\end{equation}
Defining the uncertainty $\sigma_v$ as in Eq.~\eqref{sigma_fisher} as the inverse square root of $\mathcal{F}$, we end up with
\begin{equation}
    \mathbf{E}\left(\chi^2|\,{v_p}\right)\equiv\overline{\chi^2}(v_p)=\overline{\chi^2}(v_p^{true})+\left(\frac{\Delta v_p}{\sigma_v(v_p^{true})}\right)^2+\mathcal{O}\left(\Delta v_p^3\right)\ ,
\end{equation}
where we used the fact that $v_p^{true}$ is the true parameter, so that the expectation value of $\partial\chi^2/\partial v_p$ vanishes at that point because it reaches its minimum. Note that $\overline{\chi^2}(v_p^{true})=\chi^2_{min}$ is the minimum value of the expectation value of $\chi^2$. Since the true uncertainty $\sigma_v^{true}$ on the measurement of $v_p^{true}$ is given in the MLE formalism by the deviation in phase velocity that corresponds to a one-unit increase of $\overline{\chi^2}(v_p)$ from its minimum $\overline{\chi^2}(v_p^{true})=\chi^2_{min}$, $\sigma_v(v_p^{true})$ gives a very good estimate of $\sigma_v^{true}$. The only approximation we made is the Taylor expansion to the second order. We can conclude that the Fisher formalism is valid if the Taylor expansion provides a good approximation of the $\chi^2$ curvature on a scale of $\sigma_v\!\left(v_p^{ true}\right)$ around the minimum. Since we do not know the measured value of the true parameter, we instead use fiducial values of $v_p$ (treated as a parameter) and assume that this approximation is valid for all $v_p^{fid}$. 

\section{Forecast using the CPTA significance function}
\label{apx_S_fun}
It is important to note that $\chi^2$ depends on the normalization of the ORF as discussed in the previous section. Due to this normalization issue, it is still unclear whether this statistic is the best to study here. Another possibility would be to use the significance function $\mathcal{S}$, as in the CPTA~\cite{Xu_2023}, which remains unchanged under normalization and is defined by 
\begin{equation}
\mathcal{S}=\sqrt{N_{pairs}}\frac{\text{Cov}\left(\Gamma_{ij}^{obs},\Gamma(\xi_{ij},v_p)\right)}{\sqrt{\text{Var}(\Gamma_{ij}^{obs})\text{Var}\left(\Gamma(\xi_{ij},v_p)\right)}}\ ,
\end{equation}
where variances (Var) and covariance (Cov) are calculated over the pulsar pairs $\langle i,j\rangle$. Similarly to the Fisher formalism, we use the Taylor expansion of $\mathcal{S}$ around its maximum to get
\begin{equation} \mathcal{S}=\mathcal{S}_{max}+\frac{1}{2}(\Delta v_p)^2\frac{d^2\mathcal{S}}{d v_p\,^2}\bigg|_{v_p^{fid}}+\mathcal{O}\left((\Delta v_p)^3\right)\ .
\end{equation}
With further calculations and using the approximation of neglecting terms containing second derivatives of $\Gamma$, we obtain
\begin{equation}
    \frac{d^2\mathcal{S}}{d v_p\,^2}\bigg|_{v_p^{fid}}=-\mathcal{S}_{max}\times h\left(\Gamma(\xi_{ij},v_p^{fid})\ ,\frac{\partial\Gamma}{\partial v_p}(\xi_{ij},v_p^{fid})\right)\ ,
\end{equation}
where
\begin{equation}
h(X_{ij},Y_{ij})=\frac{\text{Var}\left(X_{ij}\right)\text{Var}\left(Y_{ij}\right)-\text{Cov}\left(X_{ij},Y_{ij}\right)}{\left[\text{Var}\left(X_{ij}\right)\right]^2}\ ,
\end{equation}
where the subscripts run over the pulsar pairs. The function $h$ remains positive as a consequence of the Cauchy-Schwarz inequality. The associated uncertainty corresponding to a drop of one unit from the maximum of $\mathcal{S}$ writes
\begin{equation} \sigma_v^\mathcal{S}=\sqrt{\frac{2}{\mathcal{S}_{max}\, h\left(\Gamma(\xi_{ij},v_p^{fid}),\frac{\partial\Gamma}{\partial v_p}(\xi_{ij},v_p^{fid})\right)}}\ .
\end{equation}

However, $\mathcal{S}$ can be biased by a few highly uncertain data points in the forecast, as it does not explicitly take into account the uncertainties of the measurements $\sigma_{ij}$ \footnote{In the frequentist analysis in CPTA~\cite{Xu_2023}, this is properly handled.}. Furthermore, the function $h$ is also independent of $\sigma_{ij}$, which means that the relation between $\sigma_v^{\mathcal{S}}$ and $\sigma_{ij}$ is implicitly mapped in the value of $\mathcal{S}_{\text{max}}$, rather than being explicitly related as in the statistic $\chi^2$. As a result, forecasting $\sigma_v^{\mathcal{S}}$ in the future becomes more difficult. Moreover, it is unclear whether a drop of one unit from the maximum of $\mathcal{S}$ corresponds to a 1$\sigma$ deviation, since $\mathcal{S}$ does not follow a known probability distribution near its maximum. For these reasons, For these reasons, we choose to rely on the $\chi^2$ statistics rather than on the significance function.

\newpage
\bibliography{Biblio}

\begin{thebibliography}{45}%
\makeatletter
\providecommand \@ifxundefined [1]{%
 \@ifx{#1\undefined}
}%
\providecommand \@ifnum [1]{%
 \ifnum #1\expandafter \@firstoftwo
 \else \expandafter \@secondoftwo
 \fi
}%
\providecommand \@ifx [1]{%
 \ifx #1\expandafter \@firstoftwo
 \else \expandafter \@secondoftwo
 \fi
}%
\providecommand \natexlab [1]{#1}%
\providecommand \enquote  [1]{``#1''}%
\providecommand \bibnamefont  [1]{#1}%
\providecommand \bibfnamefont [1]{#1}%
\providecommand \citenamefont [1]{#1}%
\providecommand \href@noop [0]{\@secondoftwo}%
\providecommand \href [0]{\begingroup \@sanitize@url \@href}%
\providecommand \@href[1]{\@@startlink{#1}\@@href}%
\providecommand \@@href[1]{\endgroup#1\@@endlink}%
\providecommand \@sanitize@url [0]{\catcode `\\12\catcode `\$12\catcode `\&12\catcode `\#12\catcode `\^12\catcode `\_12\catcode `\%12\relax}%
\providecommand \@@startlink[1]{}%
\providecommand \@@endlink[0]{}%
\providecommand \url  [0]{\begingroup\@sanitize@url \@url }%
\providecommand \@url [1]{\endgroup\@href {#1}{\urlprefix }}%
\providecommand \urlprefix  [0]{URL }%
\providecommand \Eprint [0]{\href }%
\providecommand \doibase [0]{http://dx.doi.org/}%
\providecommand \selectlanguage [0]{\@gobble}%
\providecommand \bibinfo  [0]{\@secondoftwo}%
\providecommand \bibfield  [0]{\@secondoftwo}%
\providecommand \translation [1]{[#1]}%
\providecommand \BibitemOpen [0]{}%
\providecommand \bibitemStop [0]{}%
\providecommand \bibitemNoStop [0]{.\EOS\space}%
\providecommand \EOS [0]{\spacefactor3000\relax}%
\providecommand \BibitemShut  [1]{\csname bibitem#1\endcsname}%
\let\auto@bib@innerbib\@empty
%</preamble>
\bibitem [{\citenamefont {Agazie}\ \emph {et~al.}(2023)\citenamefont {Agazie}, \citenamefont {Anumarlapudi}, \citenamefont {Archibald}, \citenamefont {Arzoumanian}, \citenamefont {Baker}, \citenamefont {Bécsy}, \citenamefont {Blecha}, \citenamefont {Brazier}, \citenamefont {Brook}, \citenamefont {Burke-Spolaor} \emph {et~al.}}]{Agazie_2023}%
  \BibitemOpen
  \bibfield  {author} {\bibinfo {author} {\bibfnamefont {Gabriella}\ \bibnamefont {Agazie}}, \bibinfo {author} {\bibfnamefont {Akash}\ \bibnamefont {Anumarlapudi}}, \bibinfo {author} {\bibfnamefont {Anne~M.}\ \bibnamefont {Archibald}}, \bibinfo {author} {\bibfnamefont {Zaven}\ \bibnamefont {Arzoumanian}}, \bibinfo {author} {\bibfnamefont {Paul~T.}\ \bibnamefont {Baker}}, \bibinfo {author} {\bibfnamefont {Bence}\ \bibnamefont {Bécsy}}, \bibinfo {author} {\bibfnamefont {Laura}\ \bibnamefont {Blecha}}, \bibinfo {author} {\bibfnamefont {Adam}\ \bibnamefont {Brazier}}, \bibinfo {author} {\bibfnamefont {Paul~R.}\ \bibnamefont {Brook}}, \bibinfo {author} {\bibfnamefont {Sarah}\ \bibnamefont {Burke-Spolaor}},  \emph {et~al.},\ }\bibfield  {title} {\enquote {\bibinfo {title} {The nanograv 15 yr data set: Evidence for a gravitational-wave background},}\ }\href {\doibase 10.3847/2041-8213/acdac6} {\bibfield  {journal} {\bibinfo  {journal} {The Astrophysical Journal Letters}\ }\textbf {\bibinfo {volume} {951}},\
  \bibinfo {pages} {L8} (\bibinfo {year} {2023})}\BibitemShut {NoStop}%
\bibitem [{\citenamefont {Afzal}\ \emph {et~al.}(2023)\citenamefont {Afzal}, \citenamefont {Agazie}, \citenamefont {Anumarlapudi}, \citenamefont {Archibald}, \citenamefont {Arzoumanian}, \citenamefont {Baker}, \citenamefont {Bécsy}, \citenamefont {Blanco-Pillado}, \citenamefont {Blecha}, \citenamefont {Boddy} \emph {et~al.}}]{Afzal_2023}%
  \BibitemOpen
  \bibfield  {author} {\bibinfo {author} {\bibfnamefont {Adeela}\ \bibnamefont {Afzal}}, \bibinfo {author} {\bibfnamefont {Gabriella}\ \bibnamefont {Agazie}}, \bibinfo {author} {\bibfnamefont {Akash}\ \bibnamefont {Anumarlapudi}}, \bibinfo {author} {\bibfnamefont {Anne~M.}\ \bibnamefont {Archibald}}, \bibinfo {author} {\bibfnamefont {Zaven}\ \bibnamefont {Arzoumanian}}, \bibinfo {author} {\bibfnamefont {Paul~T.}\ \bibnamefont {Baker}}, \bibinfo {author} {\bibfnamefont {Bence}\ \bibnamefont {Bécsy}}, \bibinfo {author} {\bibfnamefont {Jose~Juan}\ \bibnamefont {Blanco-Pillado}}, \bibinfo {author} {\bibfnamefont {Laura}\ \bibnamefont {Blecha}}, \bibinfo {author} {\bibfnamefont {Kimberly~K.}\ \bibnamefont {Boddy}},  \emph {et~al.},\ }\bibfield  {title} {\enquote {\bibinfo {title} {The nanograv 15 yr data set: Search for signals from new physics},}\ }\href {\doibase 10.3847/2041-8213/acdc91} {\bibfield  {journal} {\bibinfo  {journal} {The Astrophysical Journal Letters}\ }\textbf {\bibinfo {volume} {951}},\
  \bibinfo {pages} {L11} (\bibinfo {year} {2023})}\BibitemShut {NoStop}%
\bibitem [{\citenamefont {Antoniadis}\ \emph {et~al.}(2023)\citenamefont {Antoniadis}, \citenamefont {Arumugam}, \citenamefont {Arumugam}, \citenamefont {Babak}, \citenamefont {Bagchi}, \citenamefont {Bak~Nielsen}, \citenamefont {Bassa}, \citenamefont {Bathula}, \citenamefont {Berthereau}, \citenamefont {Bonetti} \emph {et~al.}}]{EPTA_2023}%
  \BibitemOpen
  \bibfield  {author} {\bibinfo {author} {\bibfnamefont {J.}~\bibnamefont {Antoniadis}}, \bibinfo {author} {\bibfnamefont {P.}~\bibnamefont {Arumugam}}, \bibinfo {author} {\bibfnamefont {S.}~\bibnamefont {Arumugam}}, \bibinfo {author} {\bibfnamefont {S.}~\bibnamefont {Babak}}, \bibinfo {author} {\bibfnamefont {M.}~\bibnamefont {Bagchi}}, \bibinfo {author} {\bibfnamefont {A.-S.}\ \bibnamefont {Bak~Nielsen}}, \bibinfo {author} {\bibfnamefont {C.~G.}\ \bibnamefont {Bassa}}, \bibinfo {author} {\bibfnamefont {A.}~\bibnamefont {Bathula}}, \bibinfo {author} {\bibfnamefont {A.}~\bibnamefont {Berthereau}}, \bibinfo {author} {\bibfnamefont {M.}~\bibnamefont {Bonetti}},  \emph {et~al.},\ }\bibfield  {title} {\enquote {\bibinfo {title} {The second data release from the european pulsar timing array: Iii. search for gravitational wave signals},}\ }\href {\doibase 10.1051/0004-6361/202346844} {\bibfield  {journal} {\bibinfo  {journal} {Astronomy \& Astrophysics}\ }\textbf {\bibinfo {volume} {678}},\ \bibinfo {pages}
  {A50} (\bibinfo {year} {2023})}\BibitemShut {NoStop}%
\bibitem [{\citenamefont {Xu}\ \emph {et~al.}(2023)\citenamefont {Xu}, \citenamefont {Chen}, \citenamefont {Guo}, \citenamefont {Jiang}, \citenamefont {Wang}, \citenamefont {Xu}, \citenamefont {Xue}, \citenamefont {Nicolas~Caballero}, \citenamefont {Yuan}, \citenamefont {Xu} \emph {et~al.}}]{Xu_2023}%
  \BibitemOpen
  \bibfield  {author} {\bibinfo {author} {\bibfnamefont {Heng}\ \bibnamefont {Xu}}, \bibinfo {author} {\bibfnamefont {Siyuan}\ \bibnamefont {Chen}}, \bibinfo {author} {\bibfnamefont {Yanjun}\ \bibnamefont {Guo}}, \bibinfo {author} {\bibfnamefont {Jinchen}\ \bibnamefont {Jiang}}, \bibinfo {author} {\bibfnamefont {Bojun}\ \bibnamefont {Wang}}, \bibinfo {author} {\bibfnamefont {Jiangwei}\ \bibnamefont {Xu}}, \bibinfo {author} {\bibfnamefont {Zihan}\ \bibnamefont {Xue}}, \bibinfo {author} {\bibfnamefont {R.}~\bibnamefont {Nicolas~Caballero}}, \bibinfo {author} {\bibfnamefont {Jianping}\ \bibnamefont {Yuan}}, \bibinfo {author} {\bibfnamefont {Yonghua}\ \bibnamefont {Xu}},  \emph {et~al.},\ }\bibfield  {title} {\enquote {\bibinfo {title} {Searching for the nano-hertz stochastic gravitational wave background with the chinese pulsar timing array data release i},}\ }\href {\doibase 10.1088/1674-4527/acdfa5} {\bibfield  {journal} {\bibinfo  {journal} {Research in Astronomy and Astrophysics}\ }\textbf {\bibinfo {volume}
  {23}},\ \bibinfo {pages} {075024} (\bibinfo {year} {2023})}\BibitemShut {NoStop}%
\bibitem [{\citenamefont {Reardon}\ \emph {et~al.}(2023)\citenamefont {Reardon}, \citenamefont {Zic}, \citenamefont {Shannon}, \citenamefont {Hobbs}, \citenamefont {Bailes}, \citenamefont {Di~Marco}, \citenamefont {Kapur}, \citenamefont {Rogers}, \citenamefont {Thrane}, \citenamefont {Askew} \emph {et~al.}}]{Reardon_2023}%
  \BibitemOpen
  \bibfield  {author} {\bibinfo {author} {\bibfnamefont {Daniel~J.}\ \bibnamefont {Reardon}}, \bibinfo {author} {\bibfnamefont {Andrew}\ \bibnamefont {Zic}}, \bibinfo {author} {\bibfnamefont {Ryan~M.}\ \bibnamefont {Shannon}}, \bibinfo {author} {\bibfnamefont {George~B.}\ \bibnamefont {Hobbs}}, \bibinfo {author} {\bibfnamefont {Matthew}\ \bibnamefont {Bailes}}, \bibinfo {author} {\bibfnamefont {Valentina}\ \bibnamefont {Di~Marco}}, \bibinfo {author} {\bibfnamefont {Agastya}\ \bibnamefont {Kapur}}, \bibinfo {author} {\bibfnamefont {Axl~F.}\ \bibnamefont {Rogers}}, \bibinfo {author} {\bibfnamefont {Eric}\ \bibnamefont {Thrane}}, \bibinfo {author} {\bibfnamefont {Jacob}\ \bibnamefont {Askew}},  \emph {et~al.},\ }\bibfield  {title} {\enquote {\bibinfo {title} {Search for an isotropic gravitational-wave background with the parkes pulsar timing array},}\ }\href {\doibase 10.3847/2041-8213/acdd02} {\bibfield  {journal} {\bibinfo  {journal} {The Astrophysical Journal Letters}\ }\textbf {\bibinfo {volume} {951}},\
  \bibinfo {pages} {L6} (\bibinfo {year} {2023})}\BibitemShut {NoStop}%
\bibitem [{\citenamefont {Yardley}\ \emph {et~al.}(2011)\citenamefont {Yardley}, \citenamefont {Coles}, \citenamefont {Hobbs}, \citenamefont {Verbiest}, \citenamefont {Manchester}, \citenamefont {van Straten}, \citenamefont {Jenet}, \citenamefont {Bailes}, \citenamefont {Bhat}, \citenamefont {Burke-Spolaor} \emph {et~al.}}]{Yardley_2011}%
  \BibitemOpen
  \bibfield  {author} {\bibinfo {author} {\bibfnamefont {D.~R.~B.}\ \bibnamefont {Yardley}}, \bibinfo {author} {\bibfnamefont {W.~A.}\ \bibnamefont {Coles}}, \bibinfo {author} {\bibfnamefont {G.~B.}\ \bibnamefont {Hobbs}}, \bibinfo {author} {\bibfnamefont {J.~P.~W.}\ \bibnamefont {Verbiest}}, \bibinfo {author} {\bibfnamefont {R.~N.}\ \bibnamefont {Manchester}}, \bibinfo {author} {\bibfnamefont {W.}~\bibnamefont {van Straten}}, \bibinfo {author} {\bibfnamefont {F.~A.}\ \bibnamefont {Jenet}}, \bibinfo {author} {\bibfnamefont {M.}~\bibnamefont {Bailes}}, \bibinfo {author} {\bibfnamefont {N.~D.~R.}\ \bibnamefont {Bhat}}, \bibinfo {author} {\bibfnamefont {S.}~\bibnamefont {Burke-Spolaor}},  \emph {et~al.},\ }\bibfield  {title} {\enquote {\bibinfo {title} {On detection of the stochastic gravitational-wave background using the parkes pulsar timing array: On detecting a gravitational-wave background},}\ }\href {\doibase 10.1111/j.1365-2966.2011.18517.x} {\bibfield  {journal} {\bibinfo  {journal} {Monthly Notices of
  the Royal Astronomical Society}\ }\textbf {\bibinfo {volume} {414}},\ \bibinfo {pages} {1777–1787} (\bibinfo {year} {2011})}\BibitemShut {NoStop}%
\bibitem [{\citenamefont {Tarafdar}\ \emph {et~al.}(2022)\citenamefont {Tarafdar}, \citenamefont {Nobleson}, \citenamefont {Rana}, \citenamefont {Singha}, \citenamefont {Krishnakumar}, \citenamefont {Joshi}, \citenamefont {Paladi}, \citenamefont {Kolhe}, \citenamefont {Batra}, \citenamefont {Agarwal} \emph {et~al.}}]{tarafdar2022indian}%
  \BibitemOpen
  \bibfield  {author} {\bibinfo {author} {\bibfnamefont {Pratik}\ \bibnamefont {Tarafdar}}, \bibinfo {author} {\bibfnamefont {K}~\bibnamefont {Nobleson}}, \bibinfo {author} {\bibfnamefont {Prerna}\ \bibnamefont {Rana}}, \bibinfo {author} {\bibfnamefont {Jaikhomba}\ \bibnamefont {Singha}}, \bibinfo {author} {\bibfnamefont {MA}~\bibnamefont {Krishnakumar}}, \bibinfo {author} {\bibfnamefont {Bhal~Chandra}\ \bibnamefont {Joshi}}, \bibinfo {author} {\bibfnamefont {Avinash~Kumar}\ \bibnamefont {Paladi}}, \bibinfo {author} {\bibfnamefont {Neel}\ \bibnamefont {Kolhe}}, \bibinfo {author} {\bibfnamefont {Neelam~Dhanda}\ \bibnamefont {Batra}}, \bibinfo {author} {\bibfnamefont {Nikita}\ \bibnamefont {Agarwal}},  \emph {et~al.},\ }\bibfield  {title} {\enquote {\bibinfo {title} {The indian pulsar timing array: First data release},}\ }\href@noop {} {\bibfield  {journal} {\bibinfo  {journal} {Publications of the Astronomical Society of Australia}\ }\textbf {\bibinfo {volume} {39}},\ \bibinfo {pages} {e053} (\bibinfo {year}
  {2022})}\BibitemShut {NoStop}%
\bibitem [{\citenamefont {Miles}\ \emph {et~al.}(2023)\citenamefont {Miles}, \citenamefont {Shannon}, \citenamefont {Bailes}, \citenamefont {Reardon}, \citenamefont {Keith}, \citenamefont {Cameron}, \citenamefont {Parthasarathy}, \citenamefont {Shamohammadi}, \citenamefont {Spiewak}, \citenamefont {van Straten} \emph {et~al.}}]{miles2023meerkat}%
  \BibitemOpen
  \bibfield  {author} {\bibinfo {author} {\bibfnamefont {Matthew~T}\ \bibnamefont {Miles}}, \bibinfo {author} {\bibfnamefont {Ryan~M}\ \bibnamefont {Shannon}}, \bibinfo {author} {\bibfnamefont {Matthew}\ \bibnamefont {Bailes}}, \bibinfo {author} {\bibfnamefont {Daniel~J}\ \bibnamefont {Reardon}}, \bibinfo {author} {\bibfnamefont {Michael~J}\ \bibnamefont {Keith}}, \bibinfo {author} {\bibfnamefont {Andrew~D}\ \bibnamefont {Cameron}}, \bibinfo {author} {\bibfnamefont {Aditya}\ \bibnamefont {Parthasarathy}}, \bibinfo {author} {\bibfnamefont {Mohsen}\ \bibnamefont {Shamohammadi}}, \bibinfo {author} {\bibfnamefont {Renee}\ \bibnamefont {Spiewak}}, \bibinfo {author} {\bibfnamefont {Willem}\ \bibnamefont {van Straten}},  \emph {et~al.},\ }\bibfield  {title} {\enquote {\bibinfo {title} {The meerkat pulsar timing array: First data release},}\ }\href@noop {} {\bibfield  {journal} {\bibinfo  {journal} {Monthly Notices of the Royal Astronomical Society}\ }\textbf {\bibinfo {volume} {519}},\ \bibinfo {pages} {3976--3991}
  (\bibinfo {year} {2023})}\BibitemShut {NoStop}%
\bibitem [{\citenamefont {Hellings}\ and\ \citenamefont {Downs}(1983{\natexlab{a}})}]{hellings1983upper}%
  \BibitemOpen
  \bibfield  {author} {\bibinfo {author} {\bibfnamefont {RW}~\bibnamefont {Hellings}}\ and\ \bibinfo {author} {\bibfnamefont {GS}~\bibnamefont {Downs}},\ }\bibfield  {title} {\enquote {\bibinfo {title} {Upper limits on the isotropic gravitational radiation background from pulsar timing analysis},}\ }\href@noop {} {\bibfield  {journal} {\bibinfo  {journal} {Astrophysical Journal, Part 2-Letters to the Editor, vol. 265, Feb. 15, 1983, p. L39-L42.}\ }\textbf {\bibinfo {volume} {265}},\ \bibinfo {pages} {L39--L42} (\bibinfo {year} {1983}{\natexlab{a}})}\BibitemShut {NoStop}%
\bibitem [{\citenamefont {Gair}\ \emph {et~al.}(2015)\citenamefont {Gair}, \citenamefont {Romano},\ and\ \citenamefont {Taylor}}]{Gair:2015hra}%
  \BibitemOpen
  \bibfield  {author} {\bibinfo {author} {\bibfnamefont {Jonathan~R.}\ \bibnamefont {Gair}}, \bibinfo {author} {\bibfnamefont {Joseph~D.}\ \bibnamefont {Romano}}, \ and\ \bibinfo {author} {\bibfnamefont {Stephen~R.}\ \bibnamefont {Taylor}},\ }\bibfield  {title} {\enquote {\bibinfo {title} {{Mapping gravitational-wave backgrounds of arbitrary polarisation using pulsar timing arrays}},}\ }\href {\doibase 10.1103/PhysRevD.92.102003} {\bibfield  {journal} {\bibinfo  {journal} {Phys. Rev. D}\ }\textbf {\bibinfo {volume} {92}},\ \bibinfo {pages} {102003} (\bibinfo {year} {2015})},\ \Eprint {http://arxiv.org/abs/1506.08668} {arXiv:1506.08668 [gr-qc]} \BibitemShut {NoStop}%
\bibitem [{\citenamefont {Qin}\ \emph {et~al.}(2019)\citenamefont {Qin}, \citenamefont {Boddy}, \citenamefont {Kamionkowski},\ and\ \citenamefont {Dai}}]{Qin:2018yhy}%
  \BibitemOpen
  \bibfield  {author} {\bibinfo {author} {\bibfnamefont {Wenzer}\ \bibnamefont {Qin}}, \bibinfo {author} {\bibfnamefont {Kimberly~K.}\ \bibnamefont {Boddy}}, \bibinfo {author} {\bibfnamefont {Marc}\ \bibnamefont {Kamionkowski}}, \ and\ \bibinfo {author} {\bibfnamefont {Liang}\ \bibnamefont {Dai}},\ }\bibfield  {title} {\enquote {\bibinfo {title} {{Pulsar-timing arrays, astrometry, and gravitational waves}},}\ }\href {\doibase 10.1103/PhysRevD.99.063002} {\bibfield  {journal} {\bibinfo  {journal} {Phys. Rev. D}\ }\textbf {\bibinfo {volume} {99}},\ \bibinfo {pages} {063002} (\bibinfo {year} {2019})},\ \Eprint {http://arxiv.org/abs/1810.02369} {arXiv:1810.02369 [astro-ph.CO]} \BibitemShut {NoStop}%
\bibitem [{\citenamefont {Qin}\ \emph {et~al.}(2021)\citenamefont {Qin}, \citenamefont {Boddy},\ and\ \citenamefont {Kamionkowski}}]{Qin:2020hfy}%
  \BibitemOpen
  \bibfield  {author} {\bibinfo {author} {\bibfnamefont {Wenzer}\ \bibnamefont {Qin}}, \bibinfo {author} {\bibfnamefont {Kimberly~K.}\ \bibnamefont {Boddy}}, \ and\ \bibinfo {author} {\bibfnamefont {Marc}\ \bibnamefont {Kamionkowski}},\ }\bibfield  {title} {\enquote {\bibinfo {title} {{Subluminal stochastic gravitational waves in pulsar-timing arrays and astrometry}},}\ }\href {\doibase 10.1103/PhysRevD.103.024045} {\bibfield  {journal} {\bibinfo  {journal} {Phys. Rev. D}\ }\textbf {\bibinfo {volume} {103}},\ \bibinfo {pages} {024045} (\bibinfo {year} {2021})},\ \Eprint {http://arxiv.org/abs/2007.11009} {arXiv:2007.11009 [gr-qc]} \BibitemShut {NoStop}%
\bibitem [{\citenamefont {Liang}\ and\ \citenamefont {Trodden}(2021)}]{Liang_2021}%
  \BibitemOpen
  \bibfield  {author} {\bibinfo {author} {\bibfnamefont {Qiuyue}\ \bibnamefont {Liang}}\ and\ \bibinfo {author} {\bibfnamefont {Mark}\ \bibnamefont {Trodden}},\ }\bibfield  {title} {\enquote {\bibinfo {title} {Detecting the stochastic gravitational wave background from massive gravity with pulsar timing arrays},}\ }\href {\doibase 10.1103/physrevd.104.084052} {\bibfield  {journal} {\bibinfo  {journal} {Physical Review D}\ }\textbf {\bibinfo {volume} {104}} (\bibinfo {year} {2021}),\ 10.1103/physrevd.104.084052}\BibitemShut {NoStop}%
\bibitem [{\citenamefont {Bernardo}\ and\ \citenamefont {Ng}(2024{\natexlab{a}})}]{Bernardo:2023zna}%
  \BibitemOpen
  \bibfield  {author} {\bibinfo {author} {\bibfnamefont {Reginald~Christian}\ \bibnamefont {Bernardo}}\ and\ \bibinfo {author} {\bibfnamefont {Kin-Wang}\ \bibnamefont {Ng}},\ }\bibfield  {title} {\enquote {\bibinfo {title} {{Beyond the Hellings{\textendash}Downs curve: Non-Einsteinian gravitational waves in pulsar timing array correlations}},}\ }\href {\doibase 10.1051/0004-6361/202449483} {\bibfield  {journal} {\bibinfo  {journal} {Astron. Astrophys.}\ }\textbf {\bibinfo {volume} {691}},\ \bibinfo {pages} {A126} (\bibinfo {year} {2024}{\natexlab{a}})},\ \Eprint {http://arxiv.org/abs/2310.07537} {arXiv:2310.07537 [gr-qc]} \BibitemShut {NoStop}%
\bibitem [{\citenamefont {Liang}\ \emph {et~al.}(2023)\citenamefont {Liang}, \citenamefont {Lin},\ and\ \citenamefont {Trodden}}]{Liang_2023}%
  \BibitemOpen
  \bibfield  {author} {\bibinfo {author} {\bibfnamefont {Qiuyue}\ \bibnamefont {Liang}}, \bibinfo {author} {\bibfnamefont {Meng-Xiang}\ \bibnamefont {Lin}}, \ and\ \bibinfo {author} {\bibfnamefont {Mark}\ \bibnamefont {Trodden}},\ }\bibfield  {title} {\enquote {\bibinfo {title} {A test of gravity with pulsar timing arrays},}\ }\href {\doibase 10.1088/1475-7516/2023/11/042} {\bibfield  {journal} {\bibinfo  {journal} {Journal of Cosmology and Astroparticle Physics}\ }\textbf {\bibinfo {volume} {2023}},\ \bibinfo {pages} {042} (\bibinfo {year} {2023})}\BibitemShut {NoStop}%
\bibitem [{\citenamefont {Bernardo}\ and\ \citenamefont {Ng}(2023{\natexlab{a}})}]{Bernardo:2023mxc}%
  \BibitemOpen
  \bibfield  {author} {\bibinfo {author} {\bibfnamefont {Reginald~Christian}\ \bibnamefont {Bernardo}}\ and\ \bibinfo {author} {\bibfnamefont {Kin-Wang}\ \bibnamefont {Ng}},\ }\bibfield  {title} {\enquote {\bibinfo {title} {{Constraining gravitational wave propagation using pulsar timing array correlations}},}\ }\href {\doibase 10.1103/PhysRevD.107.L101502} {\bibfield  {journal} {\bibinfo  {journal} {Phys. Rev. D}\ }\textbf {\bibinfo {volume} {107}},\ \bibinfo {pages} {L101502} (\bibinfo {year} {2023}{\natexlab{a}})},\ \Eprint {http://arxiv.org/abs/2302.11796} {arXiv:2302.11796 [gr-qc]} \BibitemShut {NoStop}%
\bibitem [{\citenamefont {Hinterbichler}(2012)}]{Hinterbichler:2011tt}%
  \BibitemOpen
  \bibfield  {author} {\bibinfo {author} {\bibfnamefont {Kurt}\ \bibnamefont {Hinterbichler}},\ }\bibfield  {title} {\enquote {\bibinfo {title} {{Theoretical Aspects of Massive Gravity}},}\ }\href {\doibase 10.1103/RevModPhys.84.671} {\bibfield  {journal} {\bibinfo  {journal} {Rev. Mod. Phys.}\ }\textbf {\bibinfo {volume} {84}},\ \bibinfo {pages} {671--710} (\bibinfo {year} {2012})},\ \Eprint {http://arxiv.org/abs/1105.3735} {arXiv:1105.3735 [hep-th]} \BibitemShut {NoStop}%
\bibitem [{\citenamefont {de~Rham}\ \emph {et~al.}(2013)\citenamefont {de~Rham}, \citenamefont {Tolley},\ and\ \citenamefont {Wesley}}]{deRham:2012fw}%
  \BibitemOpen
  \bibfield  {author} {\bibinfo {author} {\bibfnamefont {Claudia}\ \bibnamefont {de~Rham}}, \bibinfo {author} {\bibfnamefont {Andrew~J.}\ \bibnamefont {Tolley}}, \ and\ \bibinfo {author} {\bibfnamefont {Daniel~H.}\ \bibnamefont {Wesley}},\ }\bibfield  {title} {\enquote {\bibinfo {title} {{Vainshtein Mechanism in Binary Pulsars}},}\ }\href {\doibase 10.1103/PhysRevD.87.044025} {\bibfield  {journal} {\bibinfo  {journal} {Phys. Rev. D}\ }\textbf {\bibinfo {volume} {87}},\ \bibinfo {pages} {044025} (\bibinfo {year} {2013})},\ \Eprint {http://arxiv.org/abs/1208.0580} {arXiv:1208.0580 [gr-qc]} \BibitemShut {NoStop}%
\bibitem [{\citenamefont {Joyce}\ \emph {et~al.}(2015)\citenamefont {Joyce}, \citenamefont {Jain}, \citenamefont {Khoury},\ and\ \citenamefont {Trodden}}]{Joyce:2014kja}%
  \BibitemOpen
  \bibfield  {author} {\bibinfo {author} {\bibfnamefont {Austin}\ \bibnamefont {Joyce}}, \bibinfo {author} {\bibfnamefont {Bhuvnesh}\ \bibnamefont {Jain}}, \bibinfo {author} {\bibfnamefont {Justin}\ \bibnamefont {Khoury}}, \ and\ \bibinfo {author} {\bibfnamefont {Mark}\ \bibnamefont {Trodden}},\ }\bibfield  {title} {\enquote {\bibinfo {title} {{Beyond the Cosmological Standard Model}},}\ }\href {\doibase 10.1016/j.physrep.2014.12.002} {\bibfield  {journal} {\bibinfo  {journal} {Phys. Rept.}\ }\textbf {\bibinfo {volume} {568}},\ \bibinfo {pages} {1--98} (\bibinfo {year} {2015})},\ \Eprint {http://arxiv.org/abs/1407.0059} {arXiv:1407.0059 [astro-ph.CO]} \BibitemShut {NoStop}%
\bibitem [{\citenamefont {Baker}\ \emph {et~al.}(2021)\citenamefont {Baker} \emph {et~al.}}]{Baker:2019gxo}%
  \BibitemOpen
  \bibfield  {author} {\bibinfo {author} {\bibfnamefont {Tessa}\ \bibnamefont {Baker}} \emph {et~al.},\ }\bibfield  {title} {\enquote {\bibinfo {title} {{Novel Probes Project: Tests of gravity on astrophysical scales}},}\ }\href {\doibase 10.1103/RevModPhys.93.015003} {\bibfield  {journal} {\bibinfo  {journal} {Rev. Mod. Phys.}\ }\textbf {\bibinfo {volume} {93}},\ \bibinfo {pages} {015003} (\bibinfo {year} {2021})},\ \Eprint {http://arxiv.org/abs/1908.03430} {arXiv:1908.03430 [astro-ph.CO]} \BibitemShut {NoStop}%
\bibitem [{\citenamefont {Liang}\ \emph {et~al.}(2024{\natexlab{a}})\citenamefont {Liang}, \citenamefont {Obata},\ and\ \citenamefont {Sasaki}}]{liang2024testinggravityfrequencydependentoverlap}%
  \BibitemOpen
  \bibfield  {author} {\bibinfo {author} {\bibfnamefont {Qiuyue}\ \bibnamefont {Liang}}, \bibinfo {author} {\bibfnamefont {Ippei}\ \bibnamefont {Obata}}, \ and\ \bibinfo {author} {\bibfnamefont {Misao}\ \bibnamefont {Sasaki}},\ }\href {https://arxiv.org/abs/2405.11755} {\enquote {\bibinfo {title} {Testing gravity with frequency-dependent overlap reduction function in pulsar timing array},}\ } (\bibinfo {year} {2024}{\natexlab{a}}),\ \Eprint {http://arxiv.org/abs/2405.11755} {arXiv:2405.11755 [astro-ph.CO]} \BibitemShut {NoStop}%
\bibitem [{\citenamefont {Bi}\ \emph {et~al.}(2024)\citenamefont {Bi}, \citenamefont {Wu}, \citenamefont {Chen},\ and\ \citenamefont {Huang}}]{Bi:2023ewq}%
  \BibitemOpen
  \bibfield  {author} {\bibinfo {author} {\bibfnamefont {Yan-Chen}\ \bibnamefont {Bi}}, \bibinfo {author} {\bibfnamefont {Yu-Mei}\ \bibnamefont {Wu}}, \bibinfo {author} {\bibfnamefont {Zu-Cheng}\ \bibnamefont {Chen}}, \ and\ \bibinfo {author} {\bibfnamefont {Qing-Guo}\ \bibnamefont {Huang}},\ }\bibfield  {title} {\enquote {\bibinfo {title} {{Constraints on the velocity of gravitational waves from the NANOGrav 15-year data set}},}\ }\href {\doibase 10.1103/PhysRevD.109.L061101} {\bibfield  {journal} {\bibinfo  {journal} {Phys. Rev. D}\ }\textbf {\bibinfo {volume} {109}},\ \bibinfo {pages} {L061101} (\bibinfo {year} {2024})},\ \Eprint {http://arxiv.org/abs/2310.08366} {arXiv:2310.08366 [astro-ph.CO]} \BibitemShut {NoStop}%
\bibitem [{\citenamefont {Bernardo}\ and\ \citenamefont {Ng}(2024{\natexlab{b}})}]{Bernardo:2023pwt}%
  \BibitemOpen
  \bibfield  {author} {\bibinfo {author} {\bibfnamefont {Reginald~Christian}\ \bibnamefont {Bernardo}}\ and\ \bibinfo {author} {\bibfnamefont {Kin-Wang}\ \bibnamefont {Ng}},\ }\bibfield  {title} {\enquote {\bibinfo {title} {{Testing gravity with cosmic variance-limited pulsar timing array correlations}},}\ }\href {\doibase 10.1103/PhysRevD.109.L101502} {\bibfield  {journal} {\bibinfo  {journal} {Phys. Rev. D}\ }\textbf {\bibinfo {volume} {109}},\ \bibinfo {pages} {L101502} (\bibinfo {year} {2024}{\natexlab{b}})},\ \Eprint {http://arxiv.org/abs/2306.13593} {arXiv:2306.13593 [gr-qc]} \BibitemShut {NoStop}%
\bibitem [{\citenamefont {Wang}\ and\ \citenamefont {Zhao}(2024)}]{Wang_2024}%
  \BibitemOpen
  \bibfield  {author} {\bibinfo {author} {\bibfnamefont {Sai}\ \bibnamefont {Wang}}\ and\ \bibinfo {author} {\bibfnamefont {Zhi-Chao}\ \bibnamefont {Zhao}},\ }\bibfield  {title} {\enquote {\bibinfo {title} {Unveiling the graviton mass bounds through the analysis of 2023 pulsar timing array data releases},}\ }\href {\doibase 10.1103/physrevd.109.l061502} {\bibfield  {journal} {\bibinfo  {journal} {Physical Review D}\ }\textbf {\bibinfo {volume} {109}} (\bibinfo {year} {2024}),\ 10.1103/physrevd.109.l061502}\BibitemShut {NoStop}%
\bibitem [{\citenamefont {Wu}\ \emph {et~al.}(2024)\citenamefont {Wu}, \citenamefont {Chen}, \citenamefont {Bi},\ and\ \citenamefont {Huang}}]{Wu_2024}%
  \BibitemOpen
  \bibfield  {author} {\bibinfo {author} {\bibfnamefont {Yu-Mei}\ \bibnamefont {Wu}}, \bibinfo {author} {\bibfnamefont {Zu-Cheng}\ \bibnamefont {Chen}}, \bibinfo {author} {\bibfnamefont {Yan-Chen}\ \bibnamefont {Bi}}, \ and\ \bibinfo {author} {\bibfnamefont {Qing-Guo}\ \bibnamefont {Huang}},\ }\bibfield  {title} {\enquote {\bibinfo {title} {Constraining the graviton mass with the nanograv 15 year data set},}\ }\href {\doibase 10.1088/1361-6382/ad2a9b} {\bibfield  {journal} {\bibinfo  {journal} {Classical and Quantum Gravity}\ }\textbf {\bibinfo {volume} {41}},\ \bibinfo {pages} {075002} (\bibinfo {year} {2024})}\BibitemShut {NoStop}%
\bibitem [{\citenamefont {Nay}\ \emph {et~al.}(2024)\citenamefont {Nay}, \citenamefont {Boddy}, \citenamefont {Smith},\ and\ \citenamefont {Mingarelli}}]{Nay:2023pwu}%
  \BibitemOpen
  \bibfield  {author} {\bibinfo {author} {\bibfnamefont {Jonathan}\ \bibnamefont {Nay}}, \bibinfo {author} {\bibfnamefont {Kimberly~K.}\ \bibnamefont {Boddy}}, \bibinfo {author} {\bibfnamefont {Tristan~L.}\ \bibnamefont {Smith}}, \ and\ \bibinfo {author} {\bibfnamefont {Chiara M.~F.}\ \bibnamefont {Mingarelli}},\ }\bibfield  {title} {\enquote {\bibinfo {title} {{Harmonic analysis for pulsar timing arrays}},}\ }\href {\doibase 10.1103/PhysRevD.110.044062} {\bibfield  {journal} {\bibinfo  {journal} {Phys. Rev. D}\ }\textbf {\bibinfo {volume} {110}},\ \bibinfo {pages} {044062} (\bibinfo {year} {2024})},\ \Eprint {http://arxiv.org/abs/2306.06168} {arXiv:2306.06168 [gr-qc]} \BibitemShut {NoStop}%
\bibitem [{\citenamefont {Bi}\ \emph {et~al.}(2026)\citenamefont {Bi}, \citenamefont {Wu},\ and\ \citenamefont {Huang}}]{Bi:2026jeu}%
  \BibitemOpen
  \bibfield  {author} {\bibinfo {author} {\bibfnamefont {Yan-Chen}\ \bibnamefont {Bi}}, \bibinfo {author} {\bibfnamefont {Yu-Mei}\ \bibnamefont {Wu}}, \ and\ \bibinfo {author} {\bibfnamefont {Qing-Guo}\ \bibnamefont {Huang}},\ }\bibfield  {title} {\enquote {\bibinfo {title} {{Harmonic Analysis on Correlation for Gravitational-Wave Backgrounds of Arbitrary Polarization from Interfering Sources in Generic Dispersion Relation}},}\ }\href@noop {} {\  (\bibinfo {year} {2026})},\ \Eprint {http://arxiv.org/abs/2602.13621} {arXiv:2602.13621 [gr-qc]} \BibitemShut {NoStop}%
\bibitem [{\citenamefont {Anholm}\ \emph {et~al.}(2009)\citenamefont {Anholm}, \citenamefont {Ballmer}, \citenamefont {Creighton}, \citenamefont {Price},\ and\ \citenamefont {Siemens}}]{Anholm_2009}%
  \BibitemOpen
  \bibfield  {author} {\bibinfo {author} {\bibfnamefont {Melissa}\ \bibnamefont {Anholm}}, \bibinfo {author} {\bibfnamefont {Stefan}\ \bibnamefont {Ballmer}}, \bibinfo {author} {\bibfnamefont {Jolien D.~E.}\ \bibnamefont {Creighton}}, \bibinfo {author} {\bibfnamefont {Larry~R.}\ \bibnamefont {Price}}, \ and\ \bibinfo {author} {\bibfnamefont {Xavier}\ \bibnamefont {Siemens}},\ }\bibfield  {title} {\enquote {\bibinfo {title} {Optimal strategies for gravitational wave stochastic background searches in pulsar timing data},}\ }\href {\doibase 10.1103/physrevd.79.084030} {\bibfield  {journal} {\bibinfo  {journal} {Physical Review D}\ }\textbf {\bibinfo {volume} {79}} (\bibinfo {year} {2009}),\ 10.1103/physrevd.79.084030}\BibitemShut {NoStop}%
\bibitem [{\citenamefont {Gair}\ \emph {et~al.}(2014)\citenamefont {Gair}, \citenamefont {Romano}, \citenamefont {Taylor},\ and\ \citenamefont {Mingarelli}}]{Gair_2014}%
  \BibitemOpen
  \bibfield  {author} {\bibinfo {author} {\bibfnamefont {Jonathan}\ \bibnamefont {Gair}}, \bibinfo {author} {\bibfnamefont {Joseph~D.}\ \bibnamefont {Romano}}, \bibinfo {author} {\bibfnamefont {Stephen}\ \bibnamefont {Taylor}}, \ and\ \bibinfo {author} {\bibfnamefont {Chiara M.~F.}\ \bibnamefont {Mingarelli}},\ }\bibfield  {title} {\enquote {\bibinfo {title} {Mapping gravitational-wave backgrounds using methods from cmb analysis: Application to pulsar timing arrays},}\ }\href {\doibase 10.1103/physrevd.90.082001} {\bibfield  {journal} {\bibinfo  {journal} {Physical Review D}\ }\textbf {\bibinfo {volume} {90}} (\bibinfo {year} {2014}),\ 10.1103/physrevd.90.082001}\BibitemShut {NoStop}%
\bibitem [{\citenamefont {Hu}\ \emph {et~al.}(2024)\citenamefont {Hu}, \citenamefont {Liang}, \citenamefont {Lin},\ and\ \citenamefont {Trodden}}]{Hu:2024wub}%
  \BibitemOpen
  \bibfield  {author} {\bibinfo {author} {\bibfnamefont {Wayne}\ \bibnamefont {Hu}}, \bibinfo {author} {\bibfnamefont {Qiuyue}\ \bibnamefont {Liang}}, \bibinfo {author} {\bibfnamefont {Meng-Xiang}\ \bibnamefont {Lin}}, \ and\ \bibinfo {author} {\bibfnamefont {Mark}\ \bibnamefont {Trodden}},\ }\bibfield  {title} {\enquote {\bibinfo {title} {{Testing gravity with realistic gravitational waveforms in Pulsar Timing Arrays}},}\ }\href {\doibase 10.1088/1475-7516/2024/12/054} {\bibfield  {journal} {\bibinfo  {journal} {JCAP}\ }\textbf {\bibinfo {volume} {12}},\ \bibinfo {pages} {054} (\bibinfo {year} {2024})},\ \Eprint {http://arxiv.org/abs/2408.11774} {arXiv:2408.11774 [astro-ph.CO]} \BibitemShut {NoStop}%
\bibitem [{\citenamefont {Hellings}\ and\ \citenamefont {Downs}(1983{\natexlab{b}})}]{Hellings:1983fr}%
  \BibitemOpen
  \bibfield  {author} {\bibinfo {author} {\bibfnamefont {R.W.}\ \bibnamefont {Hellings}}\ and\ \bibinfo {author} {\bibfnamefont {G.S.}\ \bibnamefont {Downs}},\ }\bibfield  {title} {\enquote {\bibinfo {title} {{Upper limits on the isotropic gravitational radiation background from pulsar timing analysis}},}\ }\href {\doibase 10.1086/183954} {\bibfield  {journal} {\bibinfo  {journal} {Astrophys. J. Lett.}\ }\textbf {\bibinfo {volume} {265}},\ \bibinfo {pages} {L39--L42} (\bibinfo {year} {1983}{\natexlab{b}})}\BibitemShut {NoStop}%
\bibitem [{\citenamefont {Roebber}\ and\ \citenamefont {Holder}(2017)}]{Roebber:2016jzl}%
  \BibitemOpen
  \bibfield  {author} {\bibinfo {author} {\bibfnamefont {Elinore}\ \bibnamefont {Roebber}}\ and\ \bibinfo {author} {\bibfnamefont {Gilbert}\ \bibnamefont {Holder}},\ }\bibfield  {title} {\enquote {\bibinfo {title} {{Harmonic space analysis of pulsar timing array redshift maps}},}\ }\href {\doibase 10.3847/1538-4357/835/1/21} {\bibfield  {journal} {\bibinfo  {journal} {Astrophys. J.}\ }\textbf {\bibinfo {volume} {835}},\ \bibinfo {pages} {21} (\bibinfo {year} {2017})},\ \Eprint {http://arxiv.org/abs/1609.06758} {arXiv:1609.06758 [astro-ph.CO]} \BibitemShut {NoStop}%
\bibitem [{\citenamefont {Bernardo}\ and\ \citenamefont {Ng}(2022)}]{Bernardo:2022xzl}%
  \BibitemOpen
  \bibfield  {author} {\bibinfo {author} {\bibfnamefont {Reginald~Christian}\ \bibnamefont {Bernardo}}\ and\ \bibinfo {author} {\bibfnamefont {Kin-Wang}\ \bibnamefont {Ng}},\ }\bibfield  {title} {\enquote {\bibinfo {title} {{Pulsar and cosmic variances of pulsar timing-array correlation measurements of the stochastic gravitational wave background}},}\ }\href {\doibase 10.1088/1475-7516/2022/11/046} {\bibfield  {journal} {\bibinfo  {journal} {JCAP}\ }\textbf {\bibinfo {volume} {11}},\ \bibinfo {pages} {046} (\bibinfo {year} {2022})},\ \Eprint {http://arxiv.org/abs/2209.14834} {arXiv:2209.14834 [gr-qc]} \BibitemShut {NoStop}%
\bibitem [{\citenamefont {Liang}\ \emph {et~al.}(2024{\natexlab{b}})\citenamefont {Liang}, \citenamefont {Obata},\ and\ \citenamefont {Sasaki}}]{Liang:2024mex}%
  \BibitemOpen
  \bibfield  {author} {\bibinfo {author} {\bibfnamefont {Qiuyue}\ \bibnamefont {Liang}}, \bibinfo {author} {\bibfnamefont {Ippei}\ \bibnamefont {Obata}}, \ and\ \bibinfo {author} {\bibfnamefont {Misao}\ \bibnamefont {Sasaki}},\ }\bibfield  {title} {\enquote {\bibinfo {title} {{Testing gravity with frequency-dependent overlap reduction function in Pulsar Timing Array}},}\ }\href {\doibase 10.1088/1475-7516/2024/10/097} {\bibfield  {journal} {\bibinfo  {journal} {JCAP}\ }\textbf {\bibinfo {volume} {10}},\ \bibinfo {pages} {097} (\bibinfo {year} {2024}{\natexlab{b}})},\ \Eprint {http://arxiv.org/abs/2405.11755} {arXiv:2405.11755 [astro-ph.CO]} \BibitemShut {NoStop}%
\bibitem [{\citenamefont {Dom{\`e}nech}\ and\ \citenamefont {Tsabodimos}(2024)}]{Domenech:2024pow}%
  \BibitemOpen
  \bibfield  {author} {\bibinfo {author} {\bibfnamefont {Guillem}\ \bibnamefont {Dom{\`e}nech}}\ and\ \bibinfo {author} {\bibfnamefont {Apostolos}\ \bibnamefont {Tsabodimos}},\ }\bibfield  {title} {\enquote {\bibinfo {title} {{Finite distance effects on the Hellings{\textendash}Downs curve in modified gravity}},}\ }\href {\doibase 10.1140/epjc/s10052-024-13418-w} {\bibfield  {journal} {\bibinfo  {journal} {Eur. Phys. J. C}\ }\textbf {\bibinfo {volume} {84}},\ \bibinfo {pages} {1005} (\bibinfo {year} {2024})},\ \bibinfo {note} {[Erratum: Eur.Phys.J.C 84, 1123 (2024)]},\ \Eprint {http://arxiv.org/abs/2407.21567} {arXiv:2407.21567 [gr-qc]} \BibitemShut {NoStop}%
\bibitem [{\citenamefont {Cordes}\ \emph {et~al.}(2025)\citenamefont {Cordes}, \citenamefont {Mitridate}, \citenamefont {Schmitz}, \citenamefont {Schröder},\ and\ \citenamefont {Wassner}}]{cordes2025overlapreductionfunctionpulsar}%
  \BibitemOpen
  \bibfield  {author} {\bibinfo {author} {\bibfnamefont {Nina}\ \bibnamefont {Cordes}}, \bibinfo {author} {\bibfnamefont {Andrea}\ \bibnamefont {Mitridate}}, \bibinfo {author} {\bibfnamefont {Kai}\ \bibnamefont {Schmitz}}, \bibinfo {author} {\bibfnamefont {Tobias}\ \bibnamefont {Schröder}}, \ and\ \bibinfo {author} {\bibfnamefont {Kim}\ \bibnamefont {Wassner}},\ }\href {https://arxiv.org/abs/2407.04464} {\enquote {\bibinfo {title} {On the overlap reduction function of pulsar timing array searches for gravitational waves in modified gravity},}\ } (\bibinfo {year} {2025}),\ \Eprint {http://arxiv.org/abs/2407.04464} {arXiv:2407.04464 [gr-qc]} \BibitemShut {NoStop}%
\bibitem [{\citenamefont {Allen}(2023)}]{Allen:2022dzg}%
  \BibitemOpen
  \bibfield  {author} {\bibinfo {author} {\bibfnamefont {Bruce}\ \bibnamefont {Allen}},\ }\bibfield  {title} {\enquote {\bibinfo {title} {{Variance of the Hellings-Downs correlation}},}\ }\href {\doibase 10.1103/PhysRevD.107.043018} {\bibfield  {journal} {\bibinfo  {journal} {Phys. Rev. D}\ }\textbf {\bibinfo {volume} {107}},\ \bibinfo {pages} {043018} (\bibinfo {year} {2023})},\ \Eprint {http://arxiv.org/abs/2205.05637} {arXiv:2205.05637 [gr-qc]} \BibitemShut {NoStop}%
\bibitem [{\citenamefont {Allen}\ and\ \citenamefont {Romano}(2025)}]{Allen:2024uqs}%
  \BibitemOpen
  \bibfield  {author} {\bibinfo {author} {\bibfnamefont {Bruce}\ \bibnamefont {Allen}}\ and\ \bibinfo {author} {\bibfnamefont {Joseph~D.}\ \bibnamefont {Romano}},\ }\bibfield  {title} {\enquote {\bibinfo {title} {{Optimal Reconstruction of the Hellings and Downs Correlation}},}\ }\href {\doibase 10.1103/PhysRevLett.134.031401} {\bibfield  {journal} {\bibinfo  {journal} {Phys. Rev. Lett.}\ }\textbf {\bibinfo {volume} {134}},\ \bibinfo {pages} {031401} (\bibinfo {year} {2025})},\ \Eprint {http://arxiv.org/abs/2407.10968} {arXiv:2407.10968 [gr-qc]} \BibitemShut {NoStop}%
\bibitem [{\citenamefont {Collaboration}(2025)}]{the_nanograv_collaboration_2025_16051178}%
  \BibitemOpen
  \bibfield  {author} {\bibinfo {author} {\bibfnamefont {The~NANOGrav}\ \bibnamefont {Collaboration}},\ }\href {\doibase 10.5281/zenodo.16051178} {\enquote {\bibinfo {title} {The nanograv 15-year data set},}\ } (\bibinfo {year} {2025})\BibitemShut {NoStop}%
\bibitem [{\citenamefont {Allen}\ and\ \citenamefont {Romano}(2023)}]{Allen_2023}%
  \BibitemOpen
  \bibfield  {author} {\bibinfo {author} {\bibfnamefont {Bruce}\ \bibnamefont {Allen}}\ and\ \bibinfo {author} {\bibfnamefont {Joseph~D.}\ \bibnamefont {Romano}},\ }\bibfield  {title} {\enquote {\bibinfo {title} {Hellings and downs correlation of an arbitrary set of pulsars},}\ }\href {\doibase 10.1103/physrevd.108.043026} {\bibfield  {journal} {\bibinfo  {journal} {Physical Review D}\ }\textbf {\bibinfo {volume} {108}} (\bibinfo {year} {2023}),\ 10.1103/physrevd.108.043026}\BibitemShut {NoStop}%
\bibitem [{\citenamefont {Babak}\ \emph {et~al.}(2024)\citenamefont {Babak}, \citenamefont {Falxa}, \citenamefont {Franciolini},\ and\ \citenamefont {Pieroni}}]{babak2024forecastingsensitivitypulsartiming}%
  \BibitemOpen
  \bibfield  {author} {\bibinfo {author} {\bibfnamefont {Stanislav}\ \bibnamefont {Babak}}, \bibinfo {author} {\bibfnamefont {Mikel}\ \bibnamefont {Falxa}}, \bibinfo {author} {\bibfnamefont {Gabriele}\ \bibnamefont {Franciolini}}, \ and\ \bibinfo {author} {\bibfnamefont {Mauro}\ \bibnamefont {Pieroni}},\ }\href {https://arxiv.org/abs/2404.02864} {\enquote {\bibinfo {title} {Forecasting the sensitivity of pulsar timing arrays to gravitational wave backgrounds},}\ } (\bibinfo {year} {2024}),\ \Eprint {http://arxiv.org/abs/2404.02864} {arXiv:2404.02864 [astro-ph.CO]} \BibitemShut {NoStop}%
\bibitem [{\citenamefont {Siemens}\ \emph {et~al.}(2013)\citenamefont {Siemens}, \citenamefont {Ellis}, \citenamefont {Jenet},\ and\ \citenamefont {Romano}}]{Siemens_2013}%
  \BibitemOpen
  \bibfield  {author} {\bibinfo {author} {\bibfnamefont {Xavier}\ \bibnamefont {Siemens}}, \bibinfo {author} {\bibfnamefont {Justin}\ \bibnamefont {Ellis}}, \bibinfo {author} {\bibfnamefont {Fredrick}\ \bibnamefont {Jenet}}, \ and\ \bibinfo {author} {\bibfnamefont {Joseph~D}\ \bibnamefont {Romano}},\ }\bibfield  {title} {\enquote {\bibinfo {title} {The stochastic background: scaling laws and time to detection for pulsar timing arrays},}\ }\href {\doibase 10.1088/0264-9381/30/22/224015} {\bibfield  {journal} {\bibinfo  {journal} {Classical and Quantum Gravity}\ }\textbf {\bibinfo {volume} {30}},\ \bibinfo {pages} {224015} (\bibinfo {year} {2013})}\BibitemShut {NoStop}%
\bibitem [{\citenamefont {Bernardo}\ and\ \citenamefont {Ng}(2023{\natexlab{b}})}]{Bernardo_2023}%
  \BibitemOpen
  \bibfield  {author} {\bibinfo {author} {\bibfnamefont {Reginald~Christian}\ \bibnamefont {Bernardo}}\ and\ \bibinfo {author} {\bibfnamefont {Kin-Wang}\ \bibnamefont {Ng}},\ }\bibfield  {title} {\enquote {\bibinfo {title} {Constraining gravitational wave propagation using pulsar timing array correlations},}\ }\href {\doibase 10.1103/physrevd.107.l101502} {\bibfield  {journal} {\bibinfo  {journal} {Physical Review D}\ }\textbf {\bibinfo {volume} {107}} (\bibinfo {year} {2023}{\natexlab{b}}),\ 10.1103/physrevd.107.l101502}\BibitemShut {NoStop}%
\bibitem [{\citenamefont {Janssen}\ \emph {et~al.}(2014)\citenamefont {Janssen}, \citenamefont {Hobbs}, \citenamefont {McLaughlin}, \citenamefont {Bassa}, \citenamefont {Deller}, \citenamefont {Kramer}, \citenamefont {Lee}, \citenamefont {Mingarelli}, \citenamefont {Rosado}, \citenamefont {Sanidas} \emph {et~al.}}]{janssen2014gravitationalwaveastronomyska}%
  \BibitemOpen
  \bibfield  {author} {\bibinfo {author} {\bibfnamefont {G.~H.}\ \bibnamefont {Janssen}}, \bibinfo {author} {\bibfnamefont {G.}~\bibnamefont {Hobbs}}, \bibinfo {author} {\bibfnamefont {M.}~\bibnamefont {McLaughlin}}, \bibinfo {author} {\bibfnamefont {C.~G.}\ \bibnamefont {Bassa}}, \bibinfo {author} {\bibfnamefont {A.~T.}\ \bibnamefont {Deller}}, \bibinfo {author} {\bibfnamefont {M.}~\bibnamefont {Kramer}}, \bibinfo {author} {\bibfnamefont {K.~J.}\ \bibnamefont {Lee}}, \bibinfo {author} {\bibfnamefont {C.~M.~F.}\ \bibnamefont {Mingarelli}}, \bibinfo {author} {\bibfnamefont {P.~A.}\ \bibnamefont {Rosado}}, \bibinfo {author} {\bibfnamefont {S.}~\bibnamefont {Sanidas}},  \emph {et~al.},\ }\href {https://arxiv.org/abs/1501.00127} {\enquote {\bibinfo {title} {Gravitational wave astronomy with the ska},}\ } (\bibinfo {year} {2014}),\ \Eprint {http://arxiv.org/abs/1501.00127} {arXiv:1501.00127 [astro-ph.IM]} \BibitemShut {NoStop}%
\bibitem [{\citenamefont {Joshi}\ \emph {et~al.}(2022)\citenamefont {Joshi}, \citenamefont {Gopakumar}, \citenamefont {Pandian}, \citenamefont {Prabu}, \citenamefont {Dey}, \citenamefont {Bagchi}, \citenamefont {Desai}, \citenamefont {Tarafdar}, \citenamefont {Rana}, \citenamefont {Maan} \emph {et~al.}}]{joshi2022nanohertz}%
  \BibitemOpen
  \bibfield  {author} {\bibinfo {author} {\bibfnamefont {Bhal~Chandra}\ \bibnamefont {Joshi}}, \bibinfo {author} {\bibfnamefont {Achamveedu}\ \bibnamefont {Gopakumar}}, \bibinfo {author} {\bibfnamefont {Arul}\ \bibnamefont {Pandian}}, \bibinfo {author} {\bibfnamefont {Thiagaraj}\ \bibnamefont {Prabu}}, \bibinfo {author} {\bibfnamefont {Lankeswar}\ \bibnamefont {Dey}}, \bibinfo {author} {\bibfnamefont {Manjari}\ \bibnamefont {Bagchi}}, \bibinfo {author} {\bibfnamefont {Shantanu}\ \bibnamefont {Desai}}, \bibinfo {author} {\bibfnamefont {Pratik}\ \bibnamefont {Tarafdar}}, \bibinfo {author} {\bibfnamefont {Prerna}\ \bibnamefont {Rana}}, \bibinfo {author} {\bibfnamefont {Yogesh}\ \bibnamefont {Maan}},  \emph {et~al.},\ }\bibfield  {title} {\enquote {\bibinfo {title} {Nanohertz gravitational wave astronomy during ska era: An inpta perspective},}\ }\href@noop {} {\bibfield  {journal} {\bibinfo  {journal} {Journal of Astrophysics and Astronomy}\ }\textbf {\bibinfo {volume} {43}},\ \bibinfo {pages} {98} (\bibinfo
  {year} {2022})}\BibitemShut {NoStop}%
\end{thebibliography}%

\end{document}